\documentclass[twocolumn,english,superscriptaddress,floatfix,prl]{revtex4-1} 
\usepackage{color}
\usepackage{amsmath}
\usepackage{amsfonts, amssymb,amsxtra}
\usepackage[]{graphicx}
\usepackage{grffile}
\usepackage{amsfonts} 
\newcommand{\bfk}{{\mathbf{k}}}
\newcommand{\bfp}{{\mathbf{p}}}
\newcommand{\bfq}{{\mathbf{q}}}
\newcommand{\bfee}{{\mathbf{E}}}
\newcommand{\bfx}{{\mathbf{x}}}

\newcommand{\av}[1]{\langle #1 \rangle}
\DeclareMathOperator{\arcsinh}{arcsinh}

\begin{document}

\title{Manipulation of a two-photon pump in superconductor -- semiconductor heterostructures }
\author{Paul Baireuther}
\affiliation{Institute for Theory of Condensed Matter, Karlsruhe Institute of Technology (KIT), 76131 Karlsruhe, Germany}
\affiliation{Instituut-Lorentz, Universiteit Leiden, P.O. Box 9506, 2300 RA Leiden, The Netherlands}
\author{Peter P. Orth}
\affiliation{Institute for Theory of Condensed Matter, Karlsruhe Institute of Technology (KIT), 76131 Karlsruhe, Germany}
\author{Ilya Vekhter}
\affiliation{Department of Physics and Astronomy, Louisiana State University, Baton
Rouge, Louisiana, 70803, USA}
\author{J\"org Schmalian}
\affiliation{Institute for Theory of Condensed Matter, Karlsruhe Institute of Technology (KIT), 76131 Karlsruhe, Germany}

\date{\today }

\begin{abstract}
We investigate the photon statistics, entanglement and squeezing of a
pn-junction sandwiched between two superconducting leads, and show that such
an electrically-driven photon pump generates correlated and entangled pairs
of photons. In particular, we demonstrate that the squeezing of the
fluctuations in the quadrature amplitudes of the emitted light can be manipulated by changing the relative phase of the order parameters of the superconductors. This reveals how macroscopic coherence of the superconducting state can be used to tailor the properties of a two-photon state.
\end{abstract}

\maketitle

The realization of solid-state photon sources that are capable of on-demand
generation of entangled photon pairs is highly desired for quantum
information processing and communication~\cite{Obrien:2009}, as well as for high precision
measurements~\cite{Kimble:1987,LaPorta:1987,Schnabel:2010}. Such two-photon processes are inherently non-classical, i.e.
they cannot be expressed naturally in terms of simple coherent states. Pairs
of entangled photons are routinely generated by parametric down conversion~\cite{PhysRevLett.61.2921, PhysRevLett.71.3893}. %PhysRevLett.57.2520}.
In this approach, a laser pumps a nonlinear optical crystal, leading to
extremely low overall conversion efficiencies from electrons pumped into the
laser to photon pairs out of the nonlinear crystal. These obstacles could be
overcome by the two-photon counterpart of a light-emitting diode (LED),
i.e., a device into which electrons are injected and which emits entangled photon pairs directly, leading to squeezed light. The overall quantum efficiency of
such a device could be close to unity. Recently, entanglement of an
electrically driven source of photon pairs was demonstrated, based on the
recombination of bi-excitons~\cite{Salter-Nature-2010}.
An alternative to bi-excitons are Cooper pairs. In both cases one expects that upon radiative recombination the entanglement of the electrons is inherited by the two-photon final state. In distinction to bi-excitons,
Cooper pairs form coherent two-electron states that scatter weakly among
each other, are only weakly affected by impurities, and, quite crucially,
can be manipulated externally, e.g., using SQUID geometries, Andreev
reflection at applied magnetic fields or electrically tunable Josephson
coupling. The proximity effect at superconductor-semiconductor junctions
was indeed demonstrated for InAs/InAlAs coupled to niobium~\cite{Takayanagi-PhysicaC-2010}. This is in accordance with the theoretical prediction~\cite{Hanamura-PhysStatSol-2002, PhysRevLett.103.187001} and observation~\cite{Suemune-JpnJournApplPhys-2006, Hayashi-ApplPhysExpress-2008, Idutsu-PhysStatSolC-2009, Suemune-ApplPhysExpress-2010} of an enhanced luminescence rate at such an interface. The exciting physics of a Josephson LED was discussed in the context of quantum dot~\cite{PhysRevLett.104.156802,0957-4484-21-27-274004, PhysRevLett.107.073901, PhysRevB.87.094511} and solid-state based devices~\cite{arXiv:1307.2248}.
\begin{figure}[b!]
  \centering
  \includegraphics[width=\linewidth]{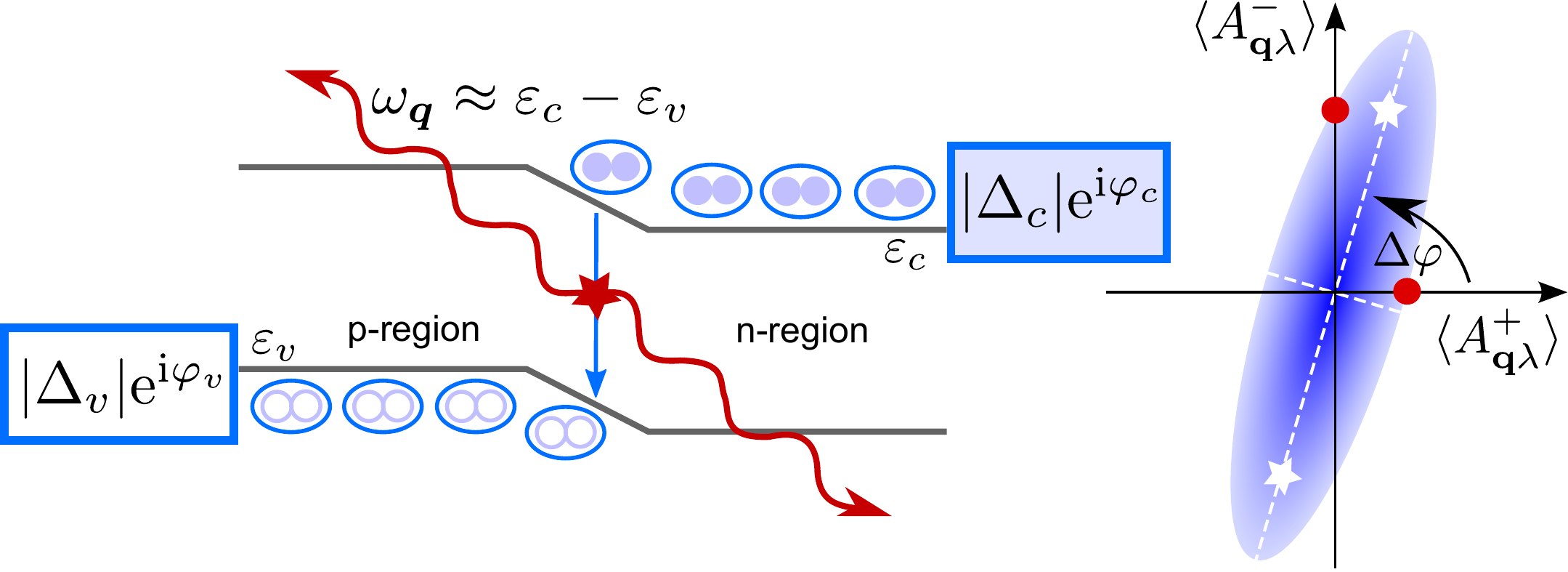}%{../Graphics/model}
  \caption{(Color online) Schematic setup of squeezed light-emitting diode consists of pn-junction with proximity induced superconductivity present in both valence $(v)$ and conduction ($c$) bands. Electronic coherence of Cooper pairs is transferred to the photons leading to two-mode squeezing of the quadrature operators $A^{\pm}_{\bfq\lambda}$ in vacuum controlled by the relative phase of the two superconductors $\Delta \varphi = \varphi_c - \varphi_v$. Entangled photon pairs of frequency of the order of the band gap $\omega_{\bfq} \approx \varepsilon_c - \varepsilon_v$ are emitted in the active region (red star). }
  \label{fig:1}
\end{figure}

In this paper we show that a superconductor-LED-superconductor
heterostructure is a source of non-classical light and, most importantly,
demonstrate how one can manipulate the two-photon coherence by varying the
relative phase between the two superconductors that are coupled to the
pn-junction. The key physical idea of our theory can be rationalized as
follows: the non-equilibrium dynamics of the photon pump can be described in
terms of an effective photon Hamiltonian that is similar to the Hamiltonian
of a quantum parametric amplifier:
\begin{equation}
\label{eq:10}
H_{\mathrm{pa}}=\omega b^{\dagger }b+\bigl( \zeta e^{-i e V_0 t} b^{\dagger }+ i\gamma e^{- i 2 e V_0 t} b^{\dagger }b^{\dagger }+ \text{h.c.} \bigr) \,,
\end{equation}%
where $b^{\dagger }$ is a photon creation operator, $\omega$ is the photon frequency (we set $\hbar =1$) and the coefficients $\zeta$ and $\gamma$ arise from pumping photons electronically via superconducting leads with applied potential difference $e V_0$. The resulting photon state of such a system is squeezed $\left\vert \psi _{%
\mathrm{photon}}(t)\right\rangle \sim \exp( \gamma t b^\dag b^\dag - \gamma^* t b b ) \left\vert
0\right\rangle $~\cite{PhysRevA.13.2226}. %Yuen1976}
 We show that the pair production amplitude $\gamma \sim |\Delta_v| |\Delta_c| e^{i \Delta \varphi}$ is determined by the gap of the two superconductors (see Fig.~\ref{fig:1}) and depends on the phase difference $\Delta \varphi $ between them. Thus, by changing the relative phase $\Delta \varphi $ of
the Cooper pair wave functions, e.g. via an external field in a SQUID
geometry, one can control the squeezing in a detuned parametric amplifier~\cite{0305-4470-17-2-031}. %entanglement of the two-photon state $%\left\vert \psi _{\mathrm{photon}}\right\rangle $.
This demonstrates how the coherence of the Cooper pair, together with the macroscopic coherence of the superconducting state can be used to tailor the properties of a two-photon
state. In what follows we substantiate this qualitative picture by explicit
many body calculations, show that the squeezing of the resulting
photon state can indeed be manipulated by changing $\Delta \varphi $ and
determine the two-photon correlation function that emerges as a result of
the superconducting coherence.

The set-up of our system is sketched in Fig.~\ref{fig:1}. We couple a pn-junction on each side to superconducting leads and apply an external voltage. The system is characterized by the Hamiltonian
\begin{equation}
H=H_{ph} +H_{c}+H_{v}+\sum_{\mathbf{k},\mathbf{q}, \sigma,\lambda}\left( g b^\dag_{%
\mathbf{q}\lambda} v_{\mathbf{k-q}\sigma }^\dag  c_{\mathbf{k}\sigma } + \text{h.c.} \right) \,, \label{H}
\end{equation}%
where $H_{ph}=\sum_{\mathbf{q},\lambda}\omega^0_{\mathbf{q}} b_{\mathbf{q}\lambda}^{\dagger }b_{\mathbf{q}\lambda}$ is the bare photon Hamiltonian. We assume emission of linearly polarized photons with $\lambda=\pm$, but the case of circular polarization, such as occurs, for example, in GaAs due to spin-orbit coupling~\cite{0957-4484-21-27-274004} is a straightforward generalization of our model. The electronic part $H_c + H_v$ 
%$H_{\alpha} = \sum_{\bfk,\sigma} \varepsilon_{\alpha, \bfk} \alpha^\dag_{\bfk, \sigma} \alpha_{\bfk,\sigma} + \sum_{\bfk} ( \Delta_\alpha \alpha^\dag_{\bfk, \uparrow} \alpha^\dag_{-\bfk, \downarrow} + \text{h.c.})$ with $\alpha \in \{c, v\}$ and $\sigma \in \{\uparrow, \downarrow\}$ 
describes the leads consisting of a superconducting conduction band on the right, with creation operators $c_{\mathbf{k}\sigma }^{\dagger }$, band dispersion $\varepsilon _{c,\mathbf{k}}$ and proximity induced superconducting gap $\Delta _{c}$, and a superconducting valence band on the left, with creation operators $v_{\mathbf{k}\sigma }^{\dagger }$, band dispersion $\varepsilon _{v,\mathbf{k}}$, and superconducting gap $\Delta_{v}$. If we include the respective chemical potentials $\mu _{c}$ and $\mu_{v}$ with $\mu _{c}-\mu _{v}=eV_0$, given by the applied voltage $V_0$, we have ($\alpha =c$ or $v$):
\begin{align}
\label{eq:11}
H_{\alpha }-\mu _{\alpha}N_{\alpha } &= \sum_{\mathbf{k}\sigma }\bigl(
\varepsilon _{\alpha \mathbf{k}}-\mu _{\alpha }\bigr) \alpha _{\mathbf{k}%
\sigma }^{\dagger }\alpha _{\mathbf{k}\sigma }  \nonumber \\
&\qquad +\sum_{\mathbf{k}}\bigl( \Delta _{\alpha }\alpha _{\mathbf{k}\uparrow
}^{\dagger }\alpha _{-\mathbf{k}\downarrow }^{\dagger }+\text{h.c.}\bigr) .
\end{align}
The coupling between photons and electrons is described by a coupling constant $g$ and leads to emission of a photon for each electron transition from conduction to valence band. We give an estimate for $g$ in the Supplemental Material. 
%The coupling between the electrons and the photons is such that each transition of an electron from the conduction to the valence band is accompanied by the emission of a photon. The coupling strength is given by the real valued coupling constant $g$.

We first derive an effective photon Hamiltonian $H_{\text{eff}}$ for this heterostructure. Technical details are provided in the Supplementary Material. Its purpose is to elucidate the nature of the photon dynamics and to obtain a tool to investigate the properties of the photon subsystem. The effective Hamiltonian is designed to generate, up to second order in perturbation theory, the same Heisenberg equations of motion for the photonic operators as the full Hamiltonian. We thus determine the equations of motion for the photon operators $b_{\mathbf{k}\sigma }^{\dagger }(t)$ perturbatively in the photon-electron coupling constant $g$ and deduce $H_{\text{eff}}$ from them. 
We deal with a nonequilibrium problem in steady state. The external bias voltage that drives the system out of equilibrium leads to time-dependent phases in the effective Hamiltonian and we obtain
\begin{align}
  \label{eq:6}
  H_{\mathrm{eff}} &=\sum_{\bfq,\lambda} \Bigl\{ \omega_{\bfq} b_{\bfq\lambda}^{\dagger } b_{\bfq\lambda} + \Bigl(g e^{-i eV_0 t} \zeta_{\bfq}(t) b_{\bfq\lambda}^\dagger \nonumber \\ & \qquad + g^2 e^{-2 i eV_0 t} \gamma_{\bfq} b_{\bfq\lambda}^{\dagger} b_{-\bfq\lambda}^{\dagger } +\text{h.c.} \Bigr) \Bigr\}\,.
\end{align}
The effective Hamiltonian $H_{\text{eff}}$ has the form of the Hamiltonian of the quantum parametric amplifier in Eq.~\eqref{eq:10} and describes electronic pumping of photons via coupling to superconducting leads. It consist of three parts that correspond to different aspects of the junction. 

The first part is the photon energy renormalized by the interaction $\omega _{\bfq}-\omega^0_{\bfq} \propto |g|^{2}$, which is of no fundamental importance to our analysis. 

\begin{figure}[t!]
  \centering
  \includegraphics[width=\linewidth]{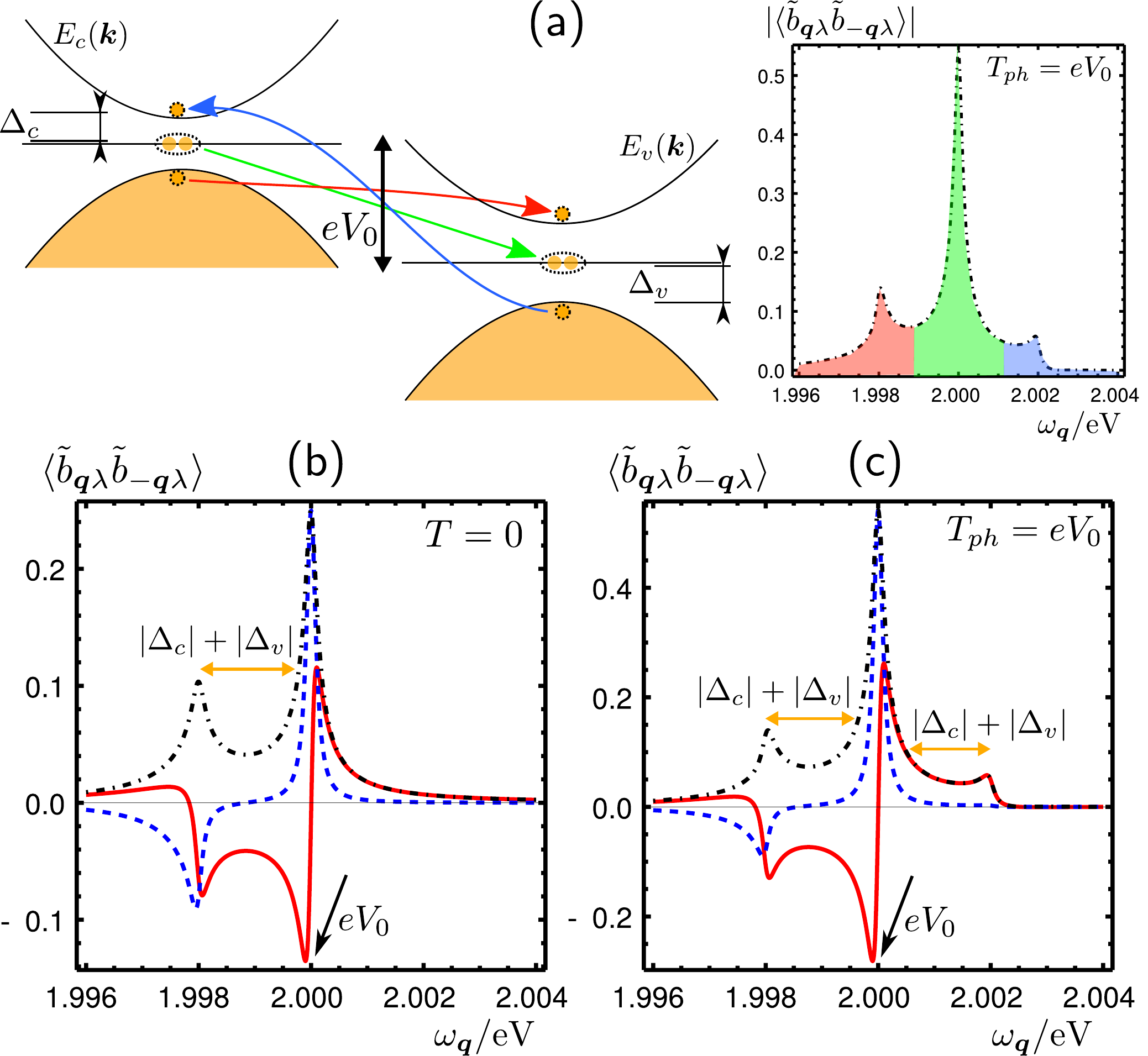}%{../Graphics/bb-FINAL-2}
  \caption{(Color online). (a) Three processes that give rise to the peaks in $|\av{\tilde b_{\bfq\lambda} \tilde b_{-\bfq\lambda}}|$ are quasi-particle tunneling from conduction to valence band and the reverse process, leading to the peaks at $\omega_\bfq = eV_0 \pm (|\Delta_c|+|\Delta_v|)$, as well as tunneling of Cooper pairs, which gives the peak at $\omega_\bfq = eV_0$. (b-c): Photon coherence $\av{\tilde b_{\bfq\lambda} \tilde b_{-\bfq\lambda}}$ at time $t = 0$ and at zero temperature (b) and finite (photon) temperature $T_{ph} = eV_0$ (c). Other parameters are chosen as $eV_0 = 2 \text{eV}$, $\Delta_c = 1 \text{meV} $, $\Delta_v = 1 \text{meV} $, $\Delta \varphi =0$, $g^2 \rho_c = \Delta_c/20$, and $\eta = 0.1 \text{meV}$. Plot shows real part (red), imaginary part (blue dashed) and absolute value (black dotted). }
  \label{fig:3}
\end{figure}

The second term describes the effect of the device being a source of single photons. It also occurs in the normal state where it describes radiation of the usual light emitting diode and produces single photons at a constant rate. In the presence of superconducting leads and the macroscopic electronic coherence, however, this term also contributes to the emergence of two-photon coherence. The different physical processes that contribute to the coherence are depicted in Fig.~\ref{fig:3}(a). The ``coefficients'' $\zeta _{\bfq}\left( t\right)$ contain fermionic creation and annihilation operators. As a consequence of the non-equilibrium state, these operators depend on the initial values of the fermionic operators well in the past where we assume that the system was decoupled and in local equilibrium. They act as random external (non-commuting) fields, which arise from the coupling of the photons to the fermionic bath, and commute with the photon operators $[b_{\mathbf{q} \lambda},\zeta_\mathbf{q}^\dagger]=0$ but not with their hermitian conjugate $[\zeta_\mathbf{q},\zeta_\mathbf{q}^\dagger] \neq 0$. %Importantly, as external fields, they do not acquire a time dependence through the effective Hamiltonian. 
At $T=0$ in the superconducting state, they are defined by correlators such as (for details consult the Supplementary Material)
\begin{align}
&\int_{-\infty}^t dt' dt'' \langle \zeta _{\bfq}(t') \zeta _{-\bfq }(t'')\rangle e^{i (\omega_\bfq-eV_0-i 0^+) (t'+t'')} \nonumber \\ &\quad = -2\sum_{\mathbf{k}}\frac{u_{c\mathbf{k}}v_{c\mathbf{k}}u_{v\mathbf{k}}^{\ast }v_{v\mathbf{k}}^{\ast } e^{2i (\omega_\bfq-eV_0)t}}{( \omega_{\bfq}-eV_0-i0^{+}) ^{2}-( E_{c\mathbf{k}}+E_{v\mathbf{k}})^{2}} \,.
\label{eta corr}
\end{align}
They contain the superconducting energy dispersions $E_{\alpha \mathbf{k}}=\sqrt{( \varepsilon _{\alpha \mathbf{k}}-\mu _{\alpha })^{2}+\vert \Delta _{\alpha }\vert ^{2}}$, the BCS coherence-factors $u_{\alpha, \mathbf k}$ and $v_{\alpha, \mathbf k}$ and we have neglected corrections to the quasi-particle energies by photon momenta $E_{\alpha \mathbf{k}+\mathbf{q}}\approx E_{\alpha \mathbf{k}}$. Most important is the third term, where at $T=0$ holds 
\begin{equation}
\label{eq:9}
  \gamma _{\bfq}=\sum_{\mathbf{k}, s=\pm }\frac{-\frac12 s \, u_{c\mathbf{k}}v_{c\mathbf{k}}u_{v\mathbf{k}}^{\ast }v_{v\mathbf{k}}^{\ast }}{(\omega_{\mathbf{q}}-eV_0-i 0^+)+s\left( E_{c\mathbf{k}}+E_{v\mathbf{k}}\right)}.
\end{equation}
This term is responsible for the fact that $H_{\mathrm{eff}}$ does have the form of a parametric amplifier, producing entangled photon pairs. Entanglement is meant in the sense that the emitted photon pairs have opposite momentum and same chirality. If the coefficients $\zeta_{\bfq}^\dagger$ and $\zeta_{\bfq}$ were numbers and would not contain fermionic operators of the initial state, we would immediately see that  Eq.~\eqref{eq:6} describes a system that produces squeezed light~\cite{PhysRevA.13.2226}. The subsequent analysis shows that this is still the case, even with the more complicated form of $H_{\mathrm{eff}}$.

To analyze whether the emitted light is squeezed we determine the uncertainties of the quadrature amplitudes of the electric field
\begin{align}
  \label{eq:5}
  \bfee(\bfx, t) &= \sum_{\bfq, \lambda} i E_{\omega_{\bfq}} \bigl( \tilde b_{\bfq\lambda} \boldsymbol{\epsilon}_{\lambda} e^{i (\bfq \cdot \bfx - \omega_\bfq t)} - \text{h.c.} \bigr)
\end{align}
with $\tilde b_{\mathbf{q}\lambda}(t)= b_{\mathbf{q}\lambda}(t) e^{i \omega_\bfq t}$, $E_{\omega_{\bfq}} = \sqrt{\omega_\bfq/2 \epsilon_0 \epsilon_r L^3}$, vacuum and relative permittivity $\epsilon_0$ and $\epsilon_r$, volume $L^3$ and linear polarization vector $\boldsymbol{\epsilon}_{\lambda}$. We consider the fluctuations of the two-mode quadrature operators
\begin{align}
  \label{eq:7}
  A_{\bfq\lambda }^{\pm}=\mathcal{N}_\pm \left[ \tilde b_{\bfq\lambda }^{\dagger } + \tilde b_{-\bfq\lambda }^{\dagger
} \pm (\tilde b_{\bfq\lambda } + \tilde b_{-\bfq\lambda }) \right]
\end{align}
with $\mathcal{N}_+ = 2^{-3/2}$ and $\mathcal{N}_- = -i 2^{-3/2}$. 
In vacuum where $\av{b^\dag_{\bfq \lambda} b_{\bfq \lambda} } = 0$, it follows directly from the bosonic commutation relations that (for details see Supplementary Material)
\begin{equation}
\Bigl\langle ( \Delta A_{\bfq\lambda }^{\pm })
^{2}\Bigr\rangle =\frac{1}{4}\bigl( 1\pm 2\text{Re}\av{\tilde b_{\mathbf{q}\lambda } \tilde b_{-\bfq\lambda }} \bigr) \,,
\end{equation}%
where $(\Delta A)^2 = (A - \av{A})^2$.
Here $\av{ \tilde b_{\bfq\lambda } \tilde b_{-\bfq\lambda}}$ depends on the superconducting phase difference $\Delta \varphi$ and can easily be
determined from our model. The key finding is that the expectation value (see Eq.~\eqref{eq:3}) that determines the uncertainties of the quadrature amplitudes of $A_{\bfq\lambda }^{\pm }$ can be changed if one changes the relative phase of the superconductor. If we picture the squeezing as an uncertainty ellipse (see Fig.~\ref{fig:1}), changing $\Delta \varphi $ simply rotates it.
\begin{figure}[t!]
  \centering
  \includegraphics[width=\linewidth]{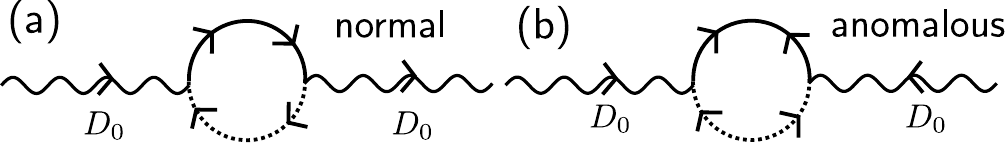}%{../Graphics/Diagrams-PhotonSelfEnergy-2}
  \caption{(a) and (b) Normal and anomalous diagrams contributing to the dressed photon propagator $D = D_0 + D_0 \Pi_{\text{el}} D_0$, where $D_0$ denotes the free photon propagator and $\Pi_{\text{el}}$ the photon self-energy  due to coupling to electrons. Solid (dashed) lines denote conduction (valence) electron propagators and wiggly lines denote photon propagators. }
  \label{fig:2}
\end{figure}

Next we address the problem of squeezing and photon statistics using a more
rigorous approach that allows more easily for a generalization to higher
order processes and feedback of the photon system onto the superconducting
leads. To this end we use the Schwinger-Keldysh formulation~\cite{Kamenev-NonEqFieldTheory-Book} of the Hamiltonian in Eq.~\eqref{H} and integrate out the fermionic degrees of freedom to arrive at an effective photonic action on the Keldysh time-contour given by
\begin{align}
  \label{eq:1}
  S^{\text{eff}}_{\text{ph}} &= \int_{-\infty}^\infty dt dt' \sum_{\bfq, \bfq',\lambda} \bar{B}_{\bfq \lambda}(t) D_{\bfq \bfq';\lambda}^{-1}(t,t') B_{\bfq' \lambda}(t') \,,
\end{align}
where $B_{\bfq \lambda}(t) = \bigl(b^{cl}_{\bfq\lambda}, \bar{b}^{cl}_{-\bfq \lambda}, b^q_{\bfq \lambda}, \bar{b}^q_{-\bfq \lambda} \bigr)^T$ carries both the Keldysh $\{cl,q\}$ and the Nambu structure. %We also include the polarization index in the momentum $\bfq \equiv (\bfq ,\lambda)$.
The photonic propagator
\begin{align}
  \label{eq:2}
  D^{-1}&= D_0^{-1} - \Pi_{\text{el}} - \Pi_{\text{bath}}
\end{align}
acquires self-energy corrections due to the coupling to the superconducting leads $\Pi_{\text{el}}$ as well as a coupling to an external (Markovian) photon bath $\Pi_{\text{bath}}$, which leads to a finite photon linewidth $\eta$~\cite{PhysRevLett.96.230602, PhysRevB.75.195331}. We consider one-loop processes such as those shown in Fig.~\ref{fig:2}(a,b). %We also consider a coupling to an external (Markovian) photon bath, which leads to a finite photon linewidth $\eta$, to arrive at
Importantly, the electronic coherence of the Cooper pairs is transferred to the photons via the anomalous elements of the photon self-energy [see Fig.~\ref{fig:2}(b)].

%For quadratic band dispersions $\varepsilon_c = \bfk^2/2 m_c^* + D/2$ and $\varepsilon_v = - \bfk^2/2m_v^* - D/2$ with effective masses $m_c^*, m_v^*$ and band  gap $D$ and first 
We first focus on the zero temperature limit in both the leads and the photonic system, and calculate the resulting photonic coherence between modes of opposite momenta by inverting the Dyson equation~\eqref{eq:2} up to second order in the electron-photon coupling $g$ to find
\begin{align}
  \label{eq:3}
  \av{\tilde b_{\bfq \lambda}(t) \tilde b_{-\bfq \lambda}(t)} &= \sum_{\bfk} \frac{2 g^2 u_{c\bfk} v_{c\bfk} u^*_{v\bfk} v^*_{v\bfk}   e^{2i (\omega_\mathbf{q}-eV_0) t} }{(\nu_\bfq -  i \eta) (\nu_\bfq + E_\bfk - i \eta)} \,.
\end{align}
%assuming real interaction constant $g$
Here, $\nu_\bfq = \omega_\bfq - e V_0$ is the photon frequency measured relative to the applied potential difference and the location of one of the resonances is determined by the Bogoliubov dispersion $E_\bfk = E_{c \bfk} + E_{v \bfk}$. The product of BCS coherence factors $u_{c\bfk} v_{c\bfk} u^*_{v\bfk} v^*_{v\bfk}= |\Delta_c| |\Delta_v| e^{i \Delta \varphi}/4 E_{c \bfk} E_{v \bfk}$ depends on the relative phase $\Delta \varphi = \varphi_c - \varphi_v$ between the two superconductors. Using the Keldysh approach, we can thus confirm the above results for the phase dependent quadrature amplitudes and our ability to manipulate light squeezing by coupling to the superconducting phase difference $\Delta \varphi$. Note that in the rotating frame of the photon fields the squeezing ellipse rotates with the detuning off the central (Cooper-pair) peak like $\av{\tilde{b}_{\bfq\lambda}(t) \tilde{b}_{-\bfq \lambda}(t)} \sim \exp[2i (\omega_\bfq-eV_0) t]$.

While we can perform the momentum sum numerically for the most general parameters, here we make a simplifying assumption of parabolic band dispersions $\varepsilon_{c\bfk} = \bfk^2/2 m_c^* + D/2$ and $\varepsilon_{v\bfk} = - \bfk^2/2m_v^* - D/2$ with effective masses $m_c^*, m_v^*$ and band gap $D$. For a symmetric choice $|\Delta_c| = |\Delta_v|\equiv |\Delta|$, $m_c^* = m_v^*$, $\mu_{c} = D/2 + \delta$ and $\mu_v = - \mu_c$, where $\delta$ defines the (quasi-) Fermi energies in the two bands related to the applied voltage by $eV_0 = D + 2 \delta$, we can evaluate the correlator analytically. The summation over momenta in Eq.~\eqref{eq:3} in this case yields
\begin{align}
  \label{eq:8}
  \av{\tilde b_{\bfq \lambda}(t) &\tilde b_{-\bfq \lambda}(t)} = g^2 \rho_c |\Delta|^2 e^{i \Delta \varphi} e^{2i (\omega_\mathbf{q}-eV_0) t}   \\ &\times \frac{ 2 \arcsin(\tilde{\nu}_\bfq - i\tilde{\eta}) + \pi [ -1 + \sqrt{1 - (\tilde{\nu}_\bfq -i \tilde{\eta})^2} ] } {(\nu_\bfq - i \eta)^2 \sqrt{ 4 |\Delta|^2 - (\nu_{\bf q} - i \eta)^2} } \nonumber \,,
\end{align}
where $\rho_c$ denotes the fermionic density of states at the Fermi surface and $\tilde{\nu}_\bfq = \nu_\bfq/2 |\Delta|$ and $\tilde{\eta} = \eta/2 |\Delta|$ are dimensionless frequencies and decay rates.

In Fig.~\ref{fig:3}(b), we present the zero temperature expectation value $\av{\tilde b_{\bfq\lambda}(t) \tilde b_{-\bfq\lambda}(t)}$ of Eq.~\eqref{eq:8} for a particular choice of parameters. The function exhibits two peaks, one at frequency $\omega_\bfq = eV_0 - |\Delta_c| - |\Delta_v|$ and one at $\omega_\bfq = eV_0$. The lower frequency peak corresponds to quasi-particle tunneling from the conduction to the valence band, while the peak at $\omega_\bfq = eV_0$ is due to tunneling of Cooper pairs. The sum over momenta only broadens the quasi-particle peak. Both processes involve the emission of photons and are thus possible at $T=0$ in the absence of a finite photon density. 

We have also obtained results for the photon coherence in the presence of a finite temperature $T_{ph}>0$ thermal background of photons. As shown in Fig.~\ref{fig:3}(c) a third peak at frequency $\omega_\bfq = eV_0 + |\Delta_c| + |\Delta_v|$ appears at photon temperature $T_{ph} > 0$, which corresponds to the absorption of photons from the thermal background and transfer of quasi-particles from the valence to the conduction band. The three processes are schematically depicted in Fig.~\ref{fig:3}(a). 

In addition we analyze the photon-photon correlation functions by coupling the photons $B_{\bfq}(t)$ to external counting fields. As expected, the density correlations between the photons of the same momentum $\av{\av{n_{\bfq\lambda} n_{\bfq\lambda}}} \equiv \av{n_{\bfq\lambda}^2} - \av{n_{\bfq\lambda}}^2$ obey the thermal relation $\av{\av{n_{\bfq\lambda} n_{\bfq\lambda} }} = \av{n_{\bfq\lambda}} [ \av{n_{\bfq\lambda}} + 1]$. These correlations simply reflect the tendency of bosonic particles to bunch. Two photons with the same momentum $\bfq$ must have been emitted in uncorrelated events since Cooper pairs have total momentum zero.

In the presence of electronic coherence, however, there also appear density correlations between photons of opposite momenta
%\begin{widetext}
\begin{align}
  \label{eq:4}
  \av{\av{n_{\bfq \lambda} n_{-\bfq \lambda}}} &= |g|^4 \sum_{\bfk, \bfk'} \frac{ |\Delta_c|^2 |\Delta_v|^2 }{ 16 E_{c\bfk} E_{c\bfk'} E_{v\bfk} E_{v\bfk'}}  \\ & \times \frac{1}{( \nu_\bfq^2 + \eta^2) ( \nu_\bfq + E_{\bfk} +i \eta) (\nu_\bfq + E_{\bfk'} - i \eta) } \nonumber\,,
\end{align}
%\end{widetext}
where $\nu_\bfq = \omega_\bfq - e V_0$ and $E_\bfk = E_{c\bfk} + E_{v \bfk}$. These correlations are inherited from the coherence of the Cooper pairs within the BCS many-body state. As before, we observe the asymmetric peak structure at zero temperatures with two peaks occurring at $\omega_\bfq = eV_0 -|\Delta_c| - |\Delta_v|$ and $\omega_\bfq= eV_0$ corresponding to tunneling of quasi-particle and Cooper pairs from the conduction to the valence band. Again, at finite (photon) temperatures a third peak emerges at $\omega_\bfq = eV_0 + |\Delta_c| + |\Delta_v|$ due to the correlated absorption of photons from the thermal background imprinting density correlations between $n_{\bfq\lambda}$ and $n_{-\bfq\lambda}$.
 
%In summary, we have shown that a pn-junction in proximity to two BCS superconductors can be operated to emit squeezed light, effects the entanglement of the emitted photons and their photon density correlations.
In summary, we have shown that a pn-junction in proximity to two BCS superconductors can be operated to emit squeezed light, produces entangled photon pairs and affects the photon density correlations. Squeezing occurs between modes of opposite momenta and results from a transfer of the electronic coherence of the Cooper pairs to the photons. The squeezing angle is controlled by the phase difference between the two superconductors. This squeezed light emitting diode enables us to use the macroscopic coherence of superconductors to manipulate the photon coherence in a two-photon pump. 
%This effect enables us to manipulate the entanglement and the density correlations of photons generated by a two-photon pump in a superconductor-semiconductor heterostructure.
\acknowledgments
We acknowledge useful discussions with M. Bechu, M. Hetterich, P. Hlobil, G. Khitrova, D. R\"ulke, M. Sch\"utt, J. Wang and M. Wegener. The Young Investigator Group of P.P.O. received financial support from the ``Concept for the Future'' of the KIT within the framework of the German Excellence Initiative. P. B. and I. V. acknowledge travel support under NSF EPSCoR Cooperative Agreement No. EPS-1003897 with additional support from the Louisiana Board of Regents. I.V. was supported in part by NSF Grant No. DMR-1105339 and ICAM Branches Cost Sharing Fund.

%\bibliography{Biblio}

\newpage
\section{Supplemental Material to ``Manipulation of a two-photon pump in superconductor -- semiconductor heterostructures'' }

\section{Effective photon Hamiltonian}
\label{sec:deriv-effect-phot}
In this section we derive the effective photon Hamiltonian $H_{\text{eff}}$ in Eq.~(4) of the main text. %We explain the individual terms and their physical meaning.
The effective Hamiltonian is an operator acting solely on the photon subspace. It generates the correct Heisenberg equations of motion for the photon operators $b_{\bfq \lambda}$ up to second order in the photon-electron coupling constant $g$.

The derivation starts from the Hamiltonian of the full system defined in Eq.~(2) of the main text
\begin{align}
  \label{eq:3}
    H &= H_{\text{ph}} + H_c + H_v + \sum_{\bfk, \bfq, \sigma, \lambda} \Bigl( g b^\dag_{\bfq \lambda} v^\dag_{\bfk - \bfq \sigma} c_{\bfk \sigma} + \text{h.c.}  \Bigr) \,,
\end{align}
where $g$ denotes the complex electron-photon coupling constant.
To obtain time-independent superconducting gap functions $\Delta_\alpha$ with $\alpha \in \{c,v\}$, one has to measure the electronic energies relative to the two \emph{different} chemical potentials $\mu_\alpha$ in the respective band, and write
\begin{align}
  \label{eq:93}
  H_{\alpha }-\mu _{\alpha}N_{\alpha } &= \sum_{\mathbf{k}\sigma }\bigl(
\varepsilon _{\alpha \mathbf{k}}-\mu _{\alpha }\bigr) \alpha _{\mathbf{k}%
\sigma }^{\dagger }\alpha _{\mathbf{k}\sigma }  \nonumber \\
&\qquad +\sum_{\mathbf{k}}\bigl( \Delta _{\alpha }\alpha _{\mathbf{k}\uparrow
}^{\dagger }\alpha _{-\mathbf{k}\downarrow }^{\dagger }+\text{h.c.}\bigr) \,.
\end{align}
%Here, $\mu_c - \mu_v = e V_0$ is determined by the applied potential difference.
The electronic Hamiltonian is diagonalized by a standard Bogoliubov transformation
\begin{align}
\label{eq:94}
 \alpha_{\bfk \sigma}
  &= u_{\alpha\bfk} \beta^\alpha_{\bfk\sigma} +\bar{\sigma} v_{\alpha\bfk} (\beta^\alpha_{-\bfk \bar{\sigma}})^\dagger \,,
\end{align}
with coherence factors $u_{\alpha \bfk} = [\frac12 ( 1+ \frac{\varepsilon_{\alpha \bfk} - \mu_\alpha}{E_{\alpha \bfk}} ) ]^{1/2} $ and $v_{\alpha \bfk} = [\frac12 ( 1- \frac{\varepsilon_{\alpha \bfk}-\mu_\alpha}{E_{\alpha \bfk}} ) ]^{1/2}$.
The total Hamiltonian in Eq.~\eqref{eq:3} transforms to
\begin{align}
  \label{eq:95}
  H &= H_{ph} + \sum_{\alpha,\bfk,\sigma} E_{\alpha\bfk} (\beta^\alpha_{\bfk\sigma})^\dagger \beta^\alpha_{\bfk\sigma}  \\ &
+ \sum_{\bfk,\bfq,\sigma} \Big( \widetilde A_{\bfq\bfk} (\beta^c_{\bfk\sigma})^\dagger \beta^v_{\bfk-\bfq\sigma} + \widetilde B_{\bfq\bfk\sigma} \beta^c_{\bfk,\sigma} \beta^v_{\bfq-\bfk\bar{\sigma}} + \text{h.c.} \Big) \nonumber
\end{align}
with $E_{\alpha \bfk} = \sqrt{(\varepsilon_{\alpha \bfk} - \mu_\alpha)^2 + |\Delta_\alpha|^2}$ and
\begin{align}
  \label{eq:96}
  \widetilde A_{\bfq\bfk} &= \bar{g} b_{\bfq \lambda} \bar{u}_{c\bfk} u_{v\bfk-\bfq} - g b_{-\bfq\lambda}^\dagger v_{c\bfk} \bar{v}_{v\bfk-\bfq} \\
\label{eq:97}
  \widetilde B_{\bfq\bfk\sigma} &= \sigma \bar{g} b_{-\bfq\lambda} \bar{v}_{c\bfk} u_{v\bfk-\bfq}  + \sigma g b_{\bfq\lambda}^\dagger u_{c\bfk} \bar{v}_{v\bfk-\bfq} \,.
\end{align}
In the Supplemental Material we denote the complex conjugate of a complex number $z$ by $\bar{z}$. The Heisenberg equations of motion for photon and Bogoliubov quasi-particle operators are given by
\begin{align}
 \label{eq:equationsOfMotion}
 \dot{b}_{H,\bfq\lambda}(t) &= i \left[H,b_{H,\bfq\lambda}(t)\right] \\
\label{eq:98}
\dot{\beta}^\alpha_{H,\bfk\sigma}(t) &= i \left[H,\beta^\alpha_{H,\bfk\sigma}(t)\right] \,.
\end{align}
In the following, we explicitly write the subscript $H$ for operators in the Heisenberg picture to distinguish them from operators in the Schr\"odinger picture that appear in the Hamiltonian.

While so far everything has been exact, we now expand the Heisenberg equations of motion up to second order of perturbation theory in the photon-electron coupling constant $g$. To do this systematically we first separate the trivial phase evolution of the photon and quasi-particle operators by defining
\begin{align}
 \label{eq:rotating_frame}
 b_{\bfq\lambda}(t) &\equiv  e^{i \omega^0_\bfq t} b_{H,\bfq\lambda}(t) \\
\label{eq:99}
\beta^\alpha_{\bfk\sigma}(t) &\equiv  e^{i E_{\alpha\bfk} t} \beta^\alpha_{H,\bfk\sigma}(t) \,.
\end{align}
In this gauge the derivatives of $b_{\bfq \lambda} (t)$ and $\beta^\alpha_{\bfk \sigma}(t)$ are at least of linear order in $g$. Formally integrating Eq.~\eqref{eq:98} yields
\begin{align}
 \label{eq:formally_integrated_QP_operators}
 \beta^\alpha_{\bfk\sigma}(t) &= i \int_{t_0}^t dt'  \bigl[H(t'), \beta^\alpha_{\bfk\sigma}(t') \bigr],
\end{align}
where $H(t)$ is the full Hamiltonian in the new gauge defined in Eqs.~\eqref{eq:rotating_frame} and~\eqref{eq:99}. We now terminate the expressions in Eq.~\eqref{eq:formally_integrated_QP_operators} after the first order in $g$, and plug them into the equation of motion of the photon operators~\eqref{eq:equationsOfMotion}. In this equation we expand in $g$ as well and neglect terms of third and higher order.

The crucial step is now the following. Since we want to find an effective photon Hamiltonian that contains only photon operators $b_{\bfq \lambda}$, we need to integrate over the fermionic degrees of freedom. Assuming that the different subsystems were decoupled and in local equilibrium in the far distant past, at time $t_0$, we know how to evaluate the expectation values of operators at that time. In particular, the two superconducting leads were in a BCS state. We thus evaluate all fermionic operators that appear as bilinears of the same band in the equation of motion~\eqref{eq:equationsOfMotion}. These operators occur in the terms that are quadratic in the coupling $g$, and we find
\begin{align}
\dot b_{\bfq\lambda}(t)
 &= -i g e^{i (\omega_\bfq-eV_0) t} \zeta_\bfq(t)
    - i |g|^2 \nu_\bfq b_{\bfq\lambda}(t_0) \nonumber \\ & \quad -i g^2 e^{2i (\omega_\bfq-eV_0) t} (\gamma_\bfq+\gamma_{-\bfq}) b_{-\bfq\lambda}^\dagger(t_0) \,.
\end{align}
Here, we have defined the complex coefficients
\begin{align}
\label{eq:104}
 \nu_\bfq
  &=-\sum_{\bfk,\sigma} \biggl(
      \frac{|u_{c\bfk}|^2 |u_{v\bfk}|^2 (n_{c\bfk\sigma}-n_{v\bfk \sigma})}{\omega_\bfq - eV_0 - E_{c\bfk}+E_{v\bfk} + i \eta} \nonumber \\ &\qquad \qquad
      - \frac{|v_{c\bfk}|^2 |v_{v\bfk}|^2 (n_{c\bfk\sigma}-n_{v\bfk \sigma})}{\omega_\bfq-eV_0+E_{c\bfk}-E_{v\bfk}+i \eta} \nonumber \\ &\qquad \qquad
      -  \frac{|u_{c\bfk}|^2 |v_{v\bfk}|^2 (1-n_{c\bfk\sigma}-n_{v\bfk \bar{\sigma}})}{\omega_\bfq-eV_0-E_{c\bfk}-E_{v\bfk}+i \eta}\nonumber \\ &\qquad \qquad
      + \frac{|v_{c\bfk}|^2 |u_{v\bfk}|^2 (1-n_{c\bfk\sigma}-n_{v\bfk\bar{\sigma}})}{\omega_\bfq-eV_0+E_{c\bfk}+E_{v\bfk}+i \eta} \biggr)
\end{align}
and
\begin{align}
\label{eq:105}
 \gamma_\bfq
  &= \frac{1}{2} \sum_{\bfk,\sigma} u_{c\bfk} v_{c \bfk} \bar{u}_{v\bfk} \bar{v}_{v,\bfk} \nonumber \\ & \quad \times \biggl(
      \frac{n_{c\bfk\sigma} - n_{v\bfk\sigma}}{\omega_\bfq-eV_0-E_{c\bfk}+E_{v\bfk}-i\eta} \nonumber \\ &
  \quad \qquad   - \frac{n_{c\bfk\sigma}-n_{v\bfk\sigma}}{\omega_\bfq-eV_0+E_{c\bfk}-E_{v\bfk}-i\eta} \nonumber \\ &
   \quad \qquad  + \frac{1-n_{c\bfk\sigma}-n_{v\bfk \bar{\sigma}}}{\omega_\bfq-eV_0-E_{c\bfk}-E_{v\bfk}-i\eta} \nonumber \\ &
  \quad \qquad  - \frac{1-n_{c\bfk\sigma}-n_{v\bfk\bar{\sigma}}}{\omega_\bfq-eV_0+E_{c\bfk}+E_{v\bfk}-i\eta} \biggr)\,,
\end{align}
where we have neglected corrections to the quasi-particle energies by photon momenta $E_{\alpha \mathbf{k}+\mathbf{q}}\approx E_{\alpha \mathbf{k}}$.
The term that appears at linear order in the coupling $g$ is given by 
\begin{align}
\label{eq:zeta}
 &\zeta_\bfq(t) \nonumber
  = \\ &-\sum_{\bfk, \sigma} \biggl(
      u_{c\bfk} \bar{u}_{v\bfk} e^{i(-E_{c\bfk}+E_{v\bfk})t} \beta^c_{\bfk\sigma, t_0} (\beta^v_{\bfk-\bfq\sigma,t_0})^\dagger \nonumber \\ & \quad \,\, 
      + v_{c\bfk} \bar{v}_{v\bfk} e^{i(E_{c\bfk}-E_{v\bfk})t} (\beta^c_{\bfk\sigma,t_0})^\dagger \beta^v_{\bfk+\bfq\sigma,t_0} \nonumber \\ & \quad \,\,
      + \sigma v_{c\bfk} \bar{u}_{v\bfk} e^{i(E_{c\bfk}+E_{v\bfk})t} (\beta^c_{\bfk\sigma,t_0})^\dagger (\beta^v_{-\bfk-\bfq\bar{\sigma},t_0})^\dagger \nonumber \\ & \quad \,\,
      - \sigma u_{c\bfk} \bar{v}_{v\bfk} e^{i(-E_{c\bfk}-E_{v\bfk})t}\beta^c_{\bfk\sigma,t_0} \beta^v_{\bfq-\bfk\bar{\sigma},t_0} \biggr)
 \,.
\end{align}
Importantly, it still contains fermionic operators $\beta^\alpha_{\bfk \sigma,t_0} \equiv \beta^\alpha_{\bfk \sigma}(t_0)$ at the initial time $t_0$. This is due to the non-equilibrium nature of the problem. Technically, it occurs since we can only trace over terms that are bilinear in fermionic operators of the same band. 
These fields do not commute with their own hermitian conjugate $[\zeta_\bfq(t), \zeta^\dag_\bfq(t)] \neq 0$.
We can work with those non-commuting coefficients $\zeta_\bfq(t)$ just as with random external fields (appearing, for example, in a Langevin equation). Since we keep terms up to $\mathcal{O}(g^2)$, we only need to know the second order correlators of the fields such as
\begin{align}
\label{eq:100}
 \langle \zeta_\bfq(t') \zeta_{-\bfq}(t'')\rangle \neq 0\,.
\end{align}
These correlators can be readily derived from the explicit expression in Eq.~\eqref{eq:zeta}. For example, at $T=0$ the correlator required to compute the anomalous photon correlations, $\av{b_{\bfq \lambda} b_{-\bfq \lambda}}$,  is given by
\begin{align}
\label{eq:101}
&\int_{-\infty}^t dt' dt'' \langle \zeta _{\bfq}(t') \zeta _{-\bfq }(t'')\rangle e^{i (\omega_\bfq-eV_0-i 0^+) (t'+t'')} \nonumber \\ &\quad = -2\sum_{\mathbf{k}}\frac{u_{c\mathbf{k}}v_{c\mathbf{k}}\bar{u}_{v\mathbf{k}} \bar{v}_{v\mathbf{k}} e^{2i (\omega_\bfq-eV_0)t}}{( \omega_{\bfq}-eV_0-i0^{+}) ^{2}-( E_{c\mathbf{k}}+E_{v\mathbf{k}})^{2}} \,.
\end{align}
Coming back to the equation of motion of the photon operators in Eq.~\eqref{eq:rotating_frame}, we observe that since the terms that contain $b_{\bfq\lambda}(t_0)$ are already of $\mathcal{O}(g^2)$, one may replace $b_{\bfq \lambda}(t_0) \rightarrow b_{\bfq\lambda}(t)$ in those terms as the difference is at least of $\mathcal{O}(g^3)$. Transforming Eq.~\eqref{eq:rotating_frame} back from the rotated to the Heisenberg frame, we finally get
\begin{align}
\label{eq:eomPhoton}
\dot{b}_{H,\bfq\lambda}(t) &= -i \omega^0_{\bfq} b_{H,\bfq\lambda}(t)
    -i g e^{-i eV_0 t} \zeta_\bfq(t) -i |g|^2 \nu_\bfq b_{H,\bfq\lambda}(t) \nonumber \\ & \quad
     -i g^2 e^{-2i eV_0 t} (\gamma_\bfq+\gamma_{-\bfq}) b_{H,-\bfq\lambda}^\dagger(t) \,.
\end{align}

%The coefficients $\nu_\bfq$ and $\gamma_\bfq$ are complex numbers, which we give below in Eqs.~\eqref{eq:104} and~\eqref{eq:105}. The coefficient of the linear term $\zeta_\bfq(t)$ on the other hand still contains fermionic operators at the initial time $t_0$. 
The effective Hamiltonian $H_{\mathrm{eff}}$ can now be read off from Eq.~\eqref{eq:eomPhoton} by demanding that it generates the correct equations of motion for the photon operators up to $\mathcal{O}(g^2)$,
\begin{align}
\label{eq:102}
 \dot b_{H,\bfq\lambda}(t) &= i \bigl[ H_{\text{eff}}, b_{H,\bfq\lambda}(t)\bigr ] + \mathcal{O}(g^3)\,,
\end{align}
and one finds
\begin{align}
\label{eq:103}
  H_{\mathrm{eff}} &=\sum_{\bfq,\lambda} \Bigl\{ \bigl(\omega^0_{\bfq} + |g|^2 \nu_\bfq \bigr) b_{\bfq\lambda}^{\dagger } b_{\bfq\lambda} + \Bigl(g e^{-i eV_0 t} \zeta_{\bfq}(t) b_{\bfq\lambda}^\dagger \nonumber \\ & \qquad + g^2 e^{-2 i eV_0 t} \gamma_{\bfq} b_{\bfq\lambda}^{\dagger} b_{-\bfq\lambda}^{\dagger } +\text{h.c.} \Bigr) \Bigr\}.
\end{align}
The first coefficient $\nu_\bfq$ is defined in Eq.~\eqref{eq:104} and leads to an unimportant renormalization of the photon frequency.
The second coefficient $\zeta_\bfq(t)$ is defined in Eq.~\eqref{eq:zeta} and appears together with a single photon operator. It has contributions which are present also in the absence of superconductivity. These terms are responsible for the photon emission in a normal light emitting diode. On the other hand, some contributions appear only in the presence of superconducting leads. Importantly, these terms also add to the anomalous photon correlator $\av{b_{\bfq \lambda} b_{-\bfq \lambda}}$. 
The last coefficient $\gamma_\bfq$ is defined in Eq.~\eqref{eq:105} and appears together with two photon operators. It is entirely due to superconductivity and thus not present in a normal light emitting diode.

Finally, we can give the $\zeta_\bfq$ and $\gamma_\bfq$ terms physical interpretations by calculating the anomalous photon correlator $\av{b_{\bfq \lambda} b_{-\bfq \lambda}}$, which is shown in Fig.~2 of the main text. At zero temperature, this expectation value exhibits two peaks: one at $\omega_\bfq = e V_0$ which corresponds to the tunneling of Cooper pairs and one at $\omega_\bfq = e V_0 - |\Delta_c| - |\Delta_v|$ which corresponds to the tunneling of quasi-particles (see Fig.~2(a)).
The $\zeta_\bfq$ term describes emission of a photon pair by splitting two \emph{different} Cooper pairs and subsequent tunneling of two quasi-particles. Due to the presence of macroscopic coherence in the superconducting state this process contributes to the anomalous photon expectation value. The $\gamma_\bfq$ term, on the other hand, also contains contributions from the tunneling of Cooper pairs.
% write sth about the interplay of the \zeta and \gamma term that gives rise to the three peak structure (at finite photon densities). with quasi-particle peaks appearing at E = ... and cooper pair peak at E = ....
% how about finite photon densities in these formulas ?
We can see this by enforcing $\gamma_\bfq =  0$ by hand, because then the (single Cooper pair) peak at $\omega_\bfq=eV_0$ vanishes. % but the peak at $\omega_\bfq = eV_0 - |\Delta_c| - |\Delta_v|$ remains. 
%In contrast, if we set $\zeta_\bfq = 0 $, both peaks remain with only quantitative differences.
% reason is that if both photons of a pair result from the same Cooper pair then their energy is higher than in the case where two Cooper pairs have to be split up.
\section{Estimate of electron-photon coupling constant $g$}
\label{sec:estim-electr-phot}
In this section, we derive an estimate of the value of the electron photon coupling constant $g$. In particular, we are interested in the product $|g|^2 \rho_c$, where $\rho_c$ is the electronic density of states, since it appears in the final formulas for the squeezing (see Eq.~(13)) and the density correlations (see Eq.~(14)). The microscopic coupling of photons and electrons is described by the dipole Hamiltonian~\cite{zubairy:qo}
\begin{align}
  \label{eq:45}
  H_{el-ph} &= \int_V d^3 x \Psi^\dag(\bfx) (- \boldsymbol{d} \cdot \boldsymbol{E}(\bfx) ) \Psi(\bfx) \,,
\end{align}
with electronic field operator $\Psi(\bfx)$ and electric dipole operator $\boldsymbol{d} = e_0 \bfx$. The electric field reads
\begin{align}
  \label{eq:46}
  \boldsymbol{E}(\bfx) = i \sum_{\bfq, \lambda} \sqrt{\frac{\hbar \omega_\bfq}{2 \epsilon_0 \epsilon_r V}} \bigl( b_{\bfq \lambda} \boldsymbol{\epsilon}_{\lambda} e^{i \bfq \bfx} - \text{h.c.} \bigr) \,,
\end{align}
where $\epsilon_0$ denotes the vacuum permittivity, $\epsilon_r$ the relative permittivity of the semiconductor and $\boldsymbol{\epsilon}_{\lambda}$ the polarization vector. We keep $\hbar$ explicitly in this section. We now expand the electronic field operator $\Psi(\bfx)$ in terms of Bloch functions $\phi_{\alpha \bfk \sigma} = e^{i \bfk \bfx} u_{\alpha \bfk}(\bfx)$ of valence and conduction band $\alpha \in \{c, v\}$ as $\Psi(\bfx) = \frac{1}{\sqrt{V}} \sum_{\alpha, \bfk, \sigma} \phi_{\alpha \bfk \sigma}(\bfx) \alpha_{\bfk \sigma}$, and estimate the relevant overlap matrix element as
\begin{align}
  \label{eq:47}
  \int \frac{d^3 x}{V} \phi^*_{c \bfk \sigma}(\bfx) \bfx \cdot \boldsymbol{\epsilon}_{\lambda} \phi_{v \bfk' \sigma}(\bfx) e^{i \bfq \bfx} \sim a_B \delta_{\bfk - \bfk' - \bfq} \delta_{\sigma \sigma'}\,,
\end{align}
where $a_B$ denotes the Bohr radius. Writing $H_{el-ph}$ in the form of Eq.~(2) yields the coupling constant
\begin{align}
  \label{eq:106}
  g &= i e_0 a_B \biggl( \frac{\hbar \omega_\bfq}{2 \epsilon_0 \epsilon_r V} \biggr)^{1/2} \,.
\end{align}
The electronic density of states is given by $\rho_c = V m_*^{3/2} \sqrt{\epsilon_F}/(\hbar^3 \sqrt{2} \pi^2)$ with effective mass $m_*$. It is convenient to express the Fermi energy in terms of the carrier density $n$. It is controlled via doping of the semiconductor as $\epsilon_F = (3 \pi^2)^{2/3} \hbar^2 n^{2/3}/(2 m_*)$, which follows from $n = k_F^3/3 \pi^2$ and $k_F^2=2 m_* \epsilon_F/\hbar^2$. The electronic density of states as a function of carrier density is thus given by 
\begin{align}
  \label{eq:107}
  \rho_c &= \frac{V m_* (3 \pi^2)^{1/3}  n^{1/3}}{2 \pi^2 \hbar^2 } \,.
\end{align}
The required product of coupling constant and density of states that enters our final result of the squeezing in Eq.~(13) is thus given by
\begin{align}
  \label{eq:108}
  |g|^2 \rho_c &= \Bigl(\frac{3}{\pi}\Bigr)^{1/3} \frac{m_*}{m_e} \frac{1}{\epsilon_r} a_B n^{1/3} \hbar \omega_\bfq \,,
\end{align}
where $m_e$ denotes the electron mass and $a_B = 0.529 \cdot 10^{-10}$ m is the Bohr radius. The numerical factor evaluates to $(3/\pi)^{1/3} = 0.985$. 

To give a numerical estimate of $|g|^2 \rho_c$ we use realistic values of GaAs, which are $m_* = 0.067 m_e$, $\hbar \omega_\bfq = 1.424$ eV, $\epsilon_r = 12.9$ and doping in the range of $n = 10^{16} \text{cm}^{-3} - 10^{19} \text{cm}^{-3}$. For this range of doping we find
\begin{align}
  \label{eq:109}
  |g|^2 \rho_c &= \, n^{1/3} \,  3.853 \cdot 10^{-11} \text{cm}\; \text{eV} \nonumber \\ 
&= 8.3 \cdot 10^{-6} \, \text{eV} - 8.3 \cdot 10^{-5} \, \text{eV} \,.
\end{align}
In the main text in Fig.~2 we use the realistic value of $|g|^2 \rho_c = 5 \cdot 10^{-5}$ eV.

\section{Two-mode squeezing}
\label{sec:anom-phot-expect}
The two-mode quadrature operators are defined in terms of the photon creation and annihilation operators as~\cite{zubairy:qo}
\begin{align}
  \label{eq:88}
  A^\pm_{\bfq \lambda} &= \mathcal{N}_{\pm} \Bigl[ \tilde{b}^\dag_{\bfq \lambda} + \tilde{b}^\dag_{-\bfq \lambda} \pm (\tilde{b}_{\bfq \lambda} + \tilde{b}_{-\bfq \lambda}) \Bigr]
\end{align}
where $\mathcal{N}_+ = 2^{-3/2}$ and $\mathcal{N}_- = - i 2^{-3/2}$. The photon operators are in the rotating frame and are related to the Heisenberg operators as $\tilde{b}_{\bfq \lambda} = b_{\bfq \lambda} e^{i \omega_\bfq t}$. The fluctuations of an operator $A$ are defined as
\begin{align}
  \label{eq:89}
  \Bigl\langle(\Delta A)^2\Bigr\rangle = \Bigl\langle ( A - \av{A})^2\Bigr\rangle ,.
\end{align}
One obtains the fluctuations of the two-mode operators straightforwardly from Eqs.~\eqref{eq:88} and~\eqref{eq:89} as
\begin{align}
  \label{eq:91}
  \Bigl\langle(\Delta A^\pm_{\bfq \lambda})^2\Bigr\rangle &= \frac14 \Bigl( 1 + \av{\tilde{b}^\dag_{\bfq \lambda} \tilde{b}_{\bfq \lambda}} + \av{\tilde{b}^\dag_{-\bfq \lambda} \tilde{b}_{-\bfq \lambda}} \nonumber \\ & \qquad \pm 2 \text{Re} \av{\tilde{b}_{\bfq \lambda} \tilde{b}_{-\bfq \lambda}} \Bigr) \,,
\end{align}
where we have used that $\av{A^\pm_{\bfq \lambda} } = 0$ and that in our system there exist no correlations between
\begin{align}
  \label{eq:90}
  \av{\tilde{b}_{\bfq \lambda} \tilde{b}_{\bfq \lambda}} = \av{\tilde{b}_{-\bfq \lambda} \tilde{b}_{-\bfq \lambda}} = \av{\tilde{b}^\dag_{-\bfq \lambda} \tilde{b}_{\bfq \lambda}} = \av{\tilde{b}^\dag_{\bfq \lambda} \tilde{b}_{-\bfq \lambda}} = 0 \,.
\end{align}
At zero temperature, the expectation values of the photon number operators vanish as well $\av{\tilde{b}^\dag_{\bfq \lambda} \tilde{b}_{\bfq \lambda}} = \av{\tilde{b}^\dag_{-\bfq \lambda} \tilde{b}_{-\bfq \lambda}} =0 $ and we are left with Eq.~(9) of the main text describing the degree of squeezing at zero temperature
\begin{align}
  \label{eq:92}
  \Bigl\langle ( \Delta A_{\bfq\lambda }^{\pm })
^{2}\Bigr\rangle =\frac{1}{4}\bigl( 1\pm 2\text{Re}\av{\tilde{b}_{\mathbf{q}\lambda } \tilde{b}_{-\bfq\lambda }} \bigr) \,.
\end{align}

% -> start from Eq.(8), definition of two-mode operators A^\pm_{q, \lambda}
% -> calculate fluctuation of this operator in Eq.(9) at finite T. Then take T=0 limit.

\section{Keldysh description of biased pn-junction with superconducting leads}
\label{sec:keldysh-descr-bias}
In this section, we describe the details of the non-equilibrium Keldysh calculation that is used to derive the anomalous expectation value $\av{\tilde{b}_{\bfq \lambda}(t) \tilde{b}_{-\bfq \lambda}(t)}$ in Eq.~(12) and the density-density correlations $\av{\av{n_{\bfq \lambda} n_{-\bfq \lambda}}}$ in Eq.~(14) of the main text. We mainly follow the notation of Ref.\cite{Kamenev-NonEqFieldTheory-Book}.

\subsection{Action on the Keldysh contour}
\label{sec:acti-keldysh-cont}
We start from the Hamiltonian of the biased pn-junction defined in Eq.~(2) of the main text
\begin{align}
  \label{eq:1}
  H &= H_{\text{ph}} + H_c + H_v + \sum_{\bfk, \bfq, \sigma, \lambda} \Bigl( g b^\dag_{\bfq \lambda} v^\dag_{\bfk - \bfq \sigma} c_{\bfk \sigma} + \text{h.c.}  \Bigr) \,.
\end{align}
Given the Hamiltonian, the Keldysh action on the closed time contour reads
\begin{align}
  \label{eq:4}
  S &=  S_{ph} + S_c + S_v - \int_{\mathcal{C}} dt \sum_{\bfk, \bfq, \sigma, \lambda} \Bigl( g \bar{b}_{\bfq \lambda} v^\dag_{\bfk - \bfq \alpha} c_{\bfk \sigma} + \text{c.c} \Bigr)
\end{align}
with $(\alpha \in \{c, v\})$
\begin{align}
  \label{eq:8}
    S_{ph} &= \int_{\mathcal{C}} dt \sum_{\bfq,\lambda} \bar{b}_{\bfq \lambda} (i \partial_t - \omega_\bfq) b_{\bfq \lambda} \\
\label{eq:9}
   S_\alpha &= \int_{\mathcal{C}} dt \sum_{\bfk \sigma} \bar{\alpha}_{\bfk \sigma} ( i \partial_t - \varepsilon_{\alpha \bfk}) \alpha_{\bfk \sigma} \nonumber \\ & \qquad \quad - \sum_{\bfk} \bigl( \Delta_\alpha \bar{\alpha}_{\bfk \uparrow} \bar{\alpha}_{-\bfk \downarrow} + \text{c.c} \bigr) \,.
\end{align}
Here, $b_{\bfq \lambda}$ denote the complex coherent state fields of the photons, $\alpha_{\bfk \sigma}$ are fermionic Grassmann fields and  $\varepsilon_{\alpha \bfk}$ describe the band dispersions.

To work with time-independent proximity induced BCS gap functions $\Delta_\alpha$ ($\alpha \in \{c, v\}$) we have to measure the electronic energies with respect to the two \emph{different} chemical potentials $\mu_\alpha$ in the two respective bands. Formally, this is achieved by a gauge transformation of the form
\begin{align}
  \label{eq:5}
  c_{\bfk\sigma} &\rightarrow \tilde{c}_{\bfk\sigma} = c_{\bfk\sigma}e^{i\mu_c t} \\
\label{eq:6}
  v_{\bfk\sigma} &\rightarrow \tilde{v}_{\bfk\sigma} = v_{\bfk\sigma}e^{i\mu_v t} \,,
\end{align}
which results in
\begin{align}
  \label{eq:2}
  S_\alpha &= \int_{\mathcal{C}} dt \sum_{\bfk, \sigma} \bigl( i \partial_t - (\epsilon_{\alpha \bfk} - \mu_\alpha) \bigr) \alpha^\dag_{\bfk \sigma} \alpha_{\bfk \sigma} \nonumber \\ & \qquad \quad - \sum_{\bfk} \bigl( \Delta_\alpha \alpha^\dag_{\bfk \uparrow} \alpha^\dag_{-\bfk \downarrow} + \text{c.c.} \bigr) \,,
\end{align}
Importantly, the coupling constant $g$ in Eq.~\eqref{eq:4} acquires a time-dependent phase factor under this gauge transformation
\begin{align}
  \label{eq:7}
  g \rightarrow g(t) = g e^{-i (\mu_c - \mu_v)t} = g e^{-i e V_0 t}
\end{align}
which rotates with the applied bias voltage $V_0$.
\subsection{Transformation from contour to RAK basis}
\label{sec:transf-from-cont}
As usual, we write the integral over the closed contour $\mathcal{C}$ as a sum of two integrals over the real line and introduce fields $b_{\bfq \lambda}^\pm$ and $\alpha_{\bfk \sigma}^\pm$ that reside on the forward ($+$) and backward ($-$) branches of the time contour~\cite{Kamenev-NonEqFieldTheory-Book}. Next, it is convenient to perform a basis change from the contour basis $(\pm)$ to the RAK (retarded-advanced-Keldysh) basis. For the complex bosonic fields, the transformation reads
\begin{align}
  \label{eq:10}
b^{cl}_{\bfq \lambda} &= \frac{1}{\sqrt{2}} \bigl( b^+_{\bfq \lambda} + b^-_{\bfq \lambda} \bigr) \\
\label{eq:11}
b^{q}_{\bfq \lambda} &= \frac{1}{\sqrt{2}} \bigl( b^+_{\bfq \lambda} - b^-_{\bfq \lambda} \bigr) \,.
\end{align}
It will be convenient to define bosonic spinors in the particle-hole channel as
\begin{align}
  \label{eq:14}
  B_{\bfq \lambda}^{cl} &= \bigl( b^{cl}_{\bfq \lambda}, \bar{b}^{cl}_{\bfq \lambda}\bigr)^T \\
\label{eq:15}
  B_{\bfq \lambda}^{q} &= \bigl( b^{q}_{\bfq \lambda}, \bar{b}^{q}_{\bfq \lambda}\bigr)^T \\
\end{align}
and define the Keldysh spinor
\begin{align}
  \label{eq:12}
  B_{\bfq \lambda} &= \bigl( B_{\bfq \lambda}^{cl}, B_{\bfq \lambda}^q \bigr)^T \,.
\end{align}
Similarly, we group the fermionic Grassmann fields into Nambu spinors
\begin{align}
  \label{eq:13}
  \Psi^1_{\bfk} &= \bigl( v_{\bfk \uparrow}^1, c_{\bfk \uparrow}^1, \bar{v}_{-\bfk \downarrow}^1, \bar{c}_{-\bfk \downarrow}^1 \bigr)^T \\
  \Psi^2_{\bfk} &= \bigl( v_{\bfk \uparrow}^2, c_{\bfk \uparrow}^2, \bar{v}_{-\bfk \downarrow}^2, \bar{c}_{-\bfk \downarrow}^2 \bigr)^T \,,
\end{align}
and define the Keldysh spinor
\begin{align}
  \label{eq:16}
  \Psi_{\bfk} &= \bigl( \Psi^1_{\bfk}, \Psi^2_{\bfk} \bigr) \,.
\end{align}
Since the fields $\alpha_{\bfk \uparrow}^j$ and $\bar{\alpha}_{-\bfk \downarrow}^j$ ($j=1,2$) are grouped together in $\Psi_{\bfk}^j$, they also transform identically under the transformation from the contour $(\pm)$ to the RAK $(1,2)$ basis
\begin{align}
  \label{eq:17}
  \alpha_{\bfk \uparrow}^1 &= \frac{1}{\sqrt{2}} ( \alpha_{\bfk \uparrow}^+ + \alpha_{\bfk \uparrow}^-) \\
\label{eq:19}
  \bar{\alpha}_{-\bfk \downarrow}^1 &= \frac{1}{\sqrt{2}} ( \bar{\alpha}_{-\bfk \downarrow}^+  + \bar{\alpha}_{\bfk \uparrow}^-) \,.
 \end{align}
The conjugates of those fields transform differently. This is possible since, unlike the bosonic case where $b_{\bfq \lambda}^\pm$ and $\bar{b}_{\bfq \lambda}^\pm$  are complex conjugates, here $\bar{\psi}_{\bfk\sigma}^\pm$ are independent Grassmann fields~\cite{Kamenev-NonEqFieldTheory-Book}. Following the Ovchinnikov-Larkin convention~\cite{LarkinOvchinnikov-Keldysh-JETP-1975}, the conjugate fields transform according to
\begin{align}
  \label{eq:18}
  \bar{\alpha}_{\bfk \uparrow}^1 &= \frac{1}{\sqrt{2}} ( \bar{\alpha}_{\bfk \uparrow}^+ - \bar{\alpha}_{\bfk \uparrow}^-) \\
\label{eq:20}
  \alpha_{-\bfk \downarrow}^1 &= \frac{1}{\sqrt{2}} ( \alpha_{-\bfk \downarrow}^+ - \alpha_{\bfk \uparrow}^-) \,.
\end{align}
The fields with $j=2$ transform according to
\begin{align}
  \label{eq:21}
  \alpha_{\bfk \uparrow}^2 &= \frac{1}{\sqrt{2}} ( \alpha_{\bfk \uparrow}^+ - \alpha_{\bfk \uparrow}^-) \\
\label{eq:22}
  \bar{\alpha}_{-\bfk \downarrow}^2 &= \frac{1}{\sqrt{2}} ( \bar{\alpha}_{-\bfk \downarrow}^+  - \bar{\alpha}_{\bfk \uparrow}^-)
\end{align}
and
\begin{align}
  \label{eq:23}
  \bar{\alpha}_{\bfk \uparrow}^2 &= \frac{1}{\sqrt{2}} ( \bar{\alpha}_{\bfk \uparrow}^+ + \bar{\alpha}_{\bfk \uparrow}^-) \\
\label{eq:24}
  \alpha_{-\bfk \downarrow}^2 &= \frac{1}{\sqrt{2}} ( \alpha_{-\bfk \downarrow}^+ + \alpha_{\bfk \uparrow}^-) \,.
\end{align}
In terms of these spinors, the action takes the form
\begin{align}
  \label{eq:25}
  &S = \int_{-\infty}^\infty dt dt' \biggl\{ \sum_{\bfq,\lambda} \bar{B}_{\bfq, \lambda} D_{0,\bfq}^{-1}(t,t') B_{\bfq, \lambda} \nonumber \\ & + \sum_{\bfk_1, \bfk_2} \bar{\Psi}_{\bfk_1}(t) \bigl[ G_{0,\bfk_1-\bfk_2}^{-1}(t,t') + V_{\bfk_1 - \bfk_2}(t, t') \bigr] \Psi_{\bfk_2}(t')\biggr\} \,.
\end{align}
In the following, we give the expressions for the free photon and electron propagators $D_{0,\bfq}(t,t')$ and $G_{0,\bfk}(t,t')$ as well as the coupling matrix $V_{\bfk}(t,t')$.
\subsection{Free photon Green's function}
\label{sec:free-photon-greens}
The free photon Green's function exhibits the block form
\begin{align}
  \label{eq:31}
  D_{0,\bfq}(t-t') &= \begin{pmatrix} D^K_{0,\bfq}(t-t')  & D^R_{0,\bfq}(t-t') \\ D^A_{0,\bfq}(t-t') & 0 \end{pmatrix}  \,,
\end{align}
where free retarded, advanced and Keldysh Green's function of the photons are defined as
\begin{align}
  \label{eq:28}
  D^{R}_{0,\bfq}(t,t') &= - i \av{B^{cl}_{\bfq\lambda}(t) \bar{B}^q_{\bfq \lambda}(t')} \\
\label{eq:29}
  D^A_{0,\bfq}(t,t') &= - i \av{B^{q}_{\bfq\lambda}(t) \bar{B}^{cl}_{\bfq \lambda}(t')} \\
\label{eq:30}
  D^K_{0,\bfq}(t,t') &= - i \av{B^{cl}_{\bfq\lambda}(t) \bar{B}^{cl}_{\bfq \lambda}(t')} \,.
\end{align}
In frequency space $D_{0,\bfq}(\omega) = \int_{-\infty}^\infty dt D_{0,\bfq}(t) e^{i \omega t}$ they are given by
\begin{align}
  \label{eq:27}
  D_{0, \bfq}^{R/A}(\omega) &= \begin{pmatrix} \frac{1}{\omega - \omega^0_\bfq \pm i 0} & 0 \\ 0 & \frac{-1}{\omega + \omega^0_\bfq \pm i 0}
\end{pmatrix} \\
D_{0, \bfq}^K(\omega) &= - 2 \pi i \bigl[2 n_B(\omega^0_\bfq) + 1 \bigr] \begin{pmatrix} \delta(\omega - \omega^0_\bfq) & 0 \\ 0 & \delta(\omega + \omega^0_\bfq) \end{pmatrix}
\end{align}
where the upper sign relates to $D^R_{0,\bfq}$. We assume inversion symmetry $\omega_{-\bfq} = \omega_\bfq$ and have introduced the Bose function $n_B(\omega_\bfq) = [ \exp(\omega_\bfq/T) - 1]^{-1}$ at temperature $T$.
\subsection{Free electronic Green's function}
\label{sec:free-electr-greens}
The free electronic Green's function exhibits the block form
\begin{align}
  \label{eq:34}
  G_{0,\bfk}(t-t') &= \begin{pmatrix} G^R_{0,\bfk}(t-t') & G^K_{0,\bfk}(t,t') \\ 0 & G^A_{0,\bfk}(t,t') \end{pmatrix} \,,
\end{align}
where the retarded, advanced and Keldysh blocks are defined as
\begin{align}
  \label{eq:26}
  G^R_{0,\bfk}(t-t') &= - i \av{\Psi^1_{\bfk}(t) \bar{\Psi}^1_\bfk(t')} \\
\label{eq:33}
  G^A_{0,\bfk}(t-t') &= - i \av{\Psi^2_{\bfk}(t) \bar{\Psi}^2_\bfk(t')} \\
\label{eq:35}
  G^K_{0,\bfk}(t-t') &= - i \av{\Psi^1_{\bfk}(t) \bar{\Psi}^2_\bfk(t')} \,.
\end{align}
Within the superconducting state the retarded and advanced blocks read in frequency space as
\begin{align}
  \label{eq:36}
  G^{R/A}_{0,\bfk}(\omega) &= \begin{pmatrix} G^{(p),R/A}_{0,v\bfk} & 0 & P^{R/A}_{0,v\bfk} & 0 \\
0 & G^{(p),R/A}_{0,c\bfk} & 0 & P^{R/A}_{0,c\bfk} \\
\bar{P}^{R/A}_{0,v\bfk} & 0 & G^{(h),R/A}_{0,v\bfk} & 0 \\
0 & \bar{P}^{R/A}_{0,c\bfk} & 0 & G^{(h),R/A}_{0,c\bfk}
\end{pmatrix}
\end{align}
where
\begin{align}
  \label{eq:38}
  G^{(p),R/A}_{0,\alpha\bfk}(\omega) &= \frac{\omega + (\epsilon_{\alpha\bfk} - \mu_v) \pm i0}{(\omega \pm i 0)^2 - E_{\alpha\bfk}^2} \\
\label{eq:39}
  G^{(h),R/A}_{0,\alpha\bfk}(\omega) &= \frac{\omega - (\epsilon_{\alpha\bfk} - \mu_c) \pm i0}{(\omega \pm i 0)^2 - E_{\alpha\bfk}^2} \\
\label{eq:40}
  P^{R/A}_{0,\alpha\bfk}(\omega) &= \frac{- \Delta_\alpha}{(\omega \pm i 0)^2 - E_{\alpha\bfk}^2} \,.
\end{align}
and the (Bogoliubov) quasi-particle energies are given by
\begin{align}
  \label{eq:37}
  E_{\alpha \bfk} &= \sqrt{(\epsilon_{\alpha\bfk} - \mu_\alpha)^2 + |\Delta_\alpha|^2} \,.
\end{align}
The Keldysh block is given by
\begin{align}
  \label{eq:41}
  &G^K_{0,\bfk}(\omega) = \nonumber \\ &= \begin{pmatrix}
G^{K}_{0,v\bfk}(\omega) & 0 & P^{K}_{0,v\bfk}(\omega) & 0 \\
0 & G^{K}_{0,c\bfk}(\omega) & 0 & P^{K}_{0,c\bfk}(\omega) \\
- \bar{P}^K_{0,v\bfk}(\omega)  & 0 & -G^K_{0,v\bfk}(-\omega) & 0 \\
0 & - \bar{P}^K_{0,c\bfk}(\omega)   & 0 & - G^K_{0,c\bfk}(-\omega)
\end{pmatrix} \,,
\end{align}
where
\begin{align}
  \label{eq:42}
  G^K_{0,\alpha\bfk}(\omega) &= - 2 \pi i \bigl( 1 - 2 n_F(E_{\alpha\bfk})\bigr) \nonumber \\ & \times \Bigl( |u_{\alpha\bfk}|^2 \delta(\omega - E_{\alpha \bfk}) - |v_{\alpha \bfk}|^2 \delta(\omega + E_{\alpha \bfk} \Bigr) \\
  P^K_{0,\alpha \bfk}(\omega) &= - 2 \pi i u_{\alpha \bfk} v_{\alpha \bfk} \bigl( 1 - 2 n_F(E_{\alpha\bfk} \bigr) \nonumber \\ & \times \Bigl(\delta(\omega - E_{\alpha \bfk}) + \delta(\omega + E_{\alpha \bfk} \Bigr) \,.
\end{align}
We have introduced the Fermi function $n_F(\omega) = [ \exp(\omega/T) + 1]^{-1}$ at temperature $T$ as well as the superconducting coherence factors $u_{\alpha \bfk} = [\frac12 ( 1+ \frac{\varepsilon_{\alpha \bfk} - \mu_\alpha}{E_{\alpha \bfk}} ) ]^{1/2} $ and $v_{\alpha \bfk} = [\frac12 ( 1- \frac{\varepsilon_{\alpha \bfk}-\mu_\alpha}{E_{\alpha \bfk}} ) ]^{1/2}$, 
$u_{\alpha \bfk} v_{\alpha \bfk} = \frac{\Delta_{\alpha \bfk}}{2 E_{\alpha \bfk}}$. 
We obtain the Green's function by expanding the electronic operators $\alpha_{\bfk\sigma}$ in terms of the Bogoliubov quasiparticle operators (see Eq.~\eqref{eq:94}). 
%as $\alpha_{\bfk \uparrow} = u_{\alpha\bfk} \gamma^\alpha_{\bfk \uparrow} - v_{\alpha\bfk} (\gamma^{\alpha}_{-\bfk \downarrow})^\dag $ and $\alpha_{-\bfk \downarrow} = u_{\alpha\bfk} \gamma^\alpha_{-\bfk \downarrow} + v_{\alpha\bfk} (\gamma^\alpha_{\bfk \uparrow})^\dag$. 
Their expectation values are then evaluated in the BCS ground state. Note that since the free electronic system is in thermal equilibrium, it holds that
\begin{align}
  \label{eq:43}
  G^K_{0,\bfk}(\omega) = F(\omega) \bigl[ G^R_{0,\bfk}(\omega) - G^A_{0,\bfk}(\omega) \bigr] \,.
\end{align}
with $F(\omega) = \tanh(\omega/2 T)$.

\subsection{Photon-electron coupling matrix}
\label{sec:phot-electr-coupl}
The photon-electron coupling matrix in Eq.~\eqref{eq:25} is given by
\begin{align}
  \label{eq:44}
  V_{\bfk}(t) &= \begin{pmatrix} V^{cl}_{\bfk}(t) & V^q_{\bfk}(t) \\ V^q_{\bfk}(t) & V^{cl}_\bfk(t)   \end{pmatrix} \,,
\end{align}
where the different blocks read ($a \in \{cl, q\}$)
\begin{align}
  \label{eq:48}
  V^a_\bfk(t) &= \frac{1}{\sqrt{2}} \begin{pmatrix}  0 & g(t) \bar{b}^a_{-\bfk} & 0 & 0 \\ \bar{g}(t) b^a_\bfk & 0 & 0 & 0 \\ 0 & 0 & 0 & - \bar{g}(t) b^a_\bfk \\ 0 & 0 & - g(t) \bar{b}^a_{-\bfk} & 0   \end{pmatrix} \,.
\end{align}

\subsection{Photon bath}
\label{sec:photon-bath}
In the following, we assume that the photons are coupled to an external photon bath. This provides a pathway for the photons to decay out of the system. Starting from the bath coupling Hamiltonian $H = \sum_{\bfq, \lambda} \omega_{\bfq}^{\text{bath}} a^\dag_{\bfq \lambda} a_{\bfq \lambda} + \sum_{\bfq, \bfp, \lambda} \lambda_{\bfq \bfp} (b^\dag_{\bfp \lambda} a_{\bfq \lambda} + \text{h.c.} )$, we follow the standard treatment of integrating over the environment~\cite{PhysRevB.75.195331}. Under the assumptions that $\lambda_{\bfq \bfp} \lambda_{\bfq \bfp'} = 0$ unless $\bfp = \bfp'$ (conservation of the in-plane photon momentum in coupling of the photon mode in two-dimensional heterostructure to bulk photon modes), that $\lambda_{\bfq \bfp} = \lambda_{\bfq}$ is independent of the system photon momentum and taking the white-noise limit of a frequency independent density of states $\rho_{\text{bath}}$ and coupling constant $\lambda$, we arrive at the bath-induced photon self-energy of the form~\cite{PhysRevB.75.195331}
\begin{align}
  \label{eq:49}
  \Pi^{R/A}_{\text{bath}}(\omega)  &= \begin{pmatrix} \mp i \eta & 0 \\ 0 & \pm i \eta \end{pmatrix} \\
\label{eq:50}
  \Pi^K_{\text{bath}}(\omega) &= 2 i \eta \begin{pmatrix} \coth \frac{\omega}{2 T} & 0 \\ 0 &  \coth \frac{-\omega}{2 T}  \end{pmatrix} \,,
\end{align}
where $\eta = \pi \lambda^2 \rho_{\text{bath}}$ and $\coth \frac{\omega}{2 T} = 2 n_B (\omega) + 1$. The photonic Green's function in Eq.~\eqref{eq:31} thus changes to
\begin{align}
  \label{eq:32}
  \tilde{D}_{0,\bfq}^{-1} = D_{0,\bfq}^{-1} - \Pi_{\text{bath}}\,.
\end{align}
This amounts to replacing in Eqs.~\eqref{eq:31}
\begin{align}
  \label{eq:51}
  \pm i 0^+ \longrightarrow \pm i \eta
\end{align}
and the delta functions in Eqs.~\eqref{eq:31} by Lorentzians
\begin{align}
  \label{eq:52}
  \delta(\omega - \omega_\bfq) \longrightarrow \frac{1}{\pi} \frac{\eta}{(\omega-\omega_\bfq)^2 + \eta^2} \,.
\end{align}

\subsection{Integration over electronic degrees of freedom}
\label{sec:integr-over-ferm}
Since we are interested in the effect of the photon-electron coupling on the properties of the photons, we choose to integrate over the fermionic degrees of freedom and analyze the resulting effective photon action $S^{\text{eff}}_{\text{ph}}$. The full action $S$ is quadratic in the fermion operators (see Eq.~\eqref{eq:25}) and the Gaussian integral yields
\begin{align}
  \label{eq:53}
  Z &= \int \mathcal{D}[B] \exp \biggl[ \int_{t,t'} \sum_\bfq i  \bar{B}_{\bfq}(t) \tilde{D}^{-1}_{0,\bfq}(t,t') B_{\bfq}(t') \nonumber \\ & + \text{Tr} \log \Bigl( - i G_0^{-1}(1 + G_0 V) \Bigr) \biggr] = \exp( i S_{\text{ph}}^{\text{eff}}) \,,
\end{align}
where $\int_{t,t'} = \int_{-\infty}^\infty dt dt'$. We write the effective photon action as
\begin{align}
  \label{eq:54}
  S_{\text{ph}}^{\text{eff}} &= \int_{-\infty}^\infty dt dt' \sum_\bfq \bar{B}_\bfq(t) D^{-1}_{\bfq}(t,t') B_{\bfq}(t')
\end{align}
with dressed photon propagator
\begin{align}
  \label{eq:55}
  D^{-1}_\bfq (t,t') &= D_{0,\bfq}^{-1}(t,t') - \Pi_{\text{bath}}(t,t') - \Pi_{\text{el},\bfq}(t,t')
\end{align}
that contains the self-energy $\Pi_{el,\bfq}(t,t')$ due to the coupling of the photons to the electronic degrees of freedom.

Expanding the $\text{Tr} \log$ to second order in the coupling constant $g$ yields the self-energy
\begin{align}
  \label{eq:57}
  \Pi_{\text{el},\bfq} (t,t') &= \begin{pmatrix} 0 & \Pi^A_{\text{el},\bfq} \\ \Pi^R_{\text{el},\bfq} & \Pi^K_{\text{el},\bfq} \end{pmatrix} \,.
\end{align}
It describes physical processes of the form depicted by the Feynman diagram in Fig.~3 of the main text. Importantly, due to the superconducting nature of the electronic system, the photonic self-energy contains non-zero anomalous contributions from diagrams of the type in Fig.~3(b) of the main text. These terms are proportional to $g(t)^2$ or $\bar{g}(t)^2$ and their phase thus evolves with the sum of the two time arguments $e^{\pm 2 i e V_0 (t + t')} $. To transform to frequency space we write the anomalous elements, which appear off-diagonal in our convention, as~\cite{Mahan-ManyParticlePhysics}
\begin{align}
  \label{eq:58}
  \Bigl( \Pi_{\text{el},\bfq}(t,t') \Bigr)_{bb/\bar{b}\bar{b}} = e^{\pm 2 i e V_0 t'} \Bigl( \tilde{\Pi}_{\text{el},\bfq}(t-t') \Bigr)_{bb/\bar{b}\bar{b}} \,,
\end{align}
where the lower sign relates to the element proportional to $g(t)^2$. The resulting functions $\Bigl( \tilde{\Pi}_{\text{el},\bfq}(t-t') \Bigr)_{bb/\bar{b}\bar{b}}$ depend only on the relative time arguments and can thus be transformed to frequency space.
Explicitly, the different components of the photonic self-energy due to the coupling to the electrons evaluate to
\begin{align}
  \label{eq:56}
  \Pi_{\text{el},\bfq}^{R/A/K}(\omega) &= \begin{pmatrix} \Bigl(\Pi_{\text{el},\bfq}^{R/A/K}(\omega)\Bigr)_{11} & \Bigl(\tilde{\Pi}_{\text{el},\bfq}^{R/A/K}(\omega)\Bigr)_{12} \\
\Bigl(\tilde{\Pi}_{\text{el},\bfq}^{R/A/K}(\omega)\Bigr)_{21} & \Bigl(\Pi_{\text{el},\bfq}^{R/A/K}(\omega)\Bigr)_{22} \,,
\end{pmatrix}
\end{align}
where the retarded and advanced components read
\begin{align}
  \label{eq:60}
  \Bigl(\Pi_{\text{el},\bfq}^{R/A}(\omega)\Bigr)_{11} &= |g|^2 \sum_{\bfk} \biggl\{ \frac{|u_c|^2 |v_v|^2}{\omega_- - E_\bfk \pm i \eta} - \frac{|u_v|^2 |v_c|^2}{\omega_- + E_\bfk \pm i \eta} \biggr\} \\
\label{eq:61}
  \Bigl(\tilde{\Pi}_{\text{el},\bfq}^{R/A}(\omega)\Bigr)_{12} &=  g^2 \sum_{\bfk} \biggl\{ \frac{u_c \bar{u}_v v_c \bar{v}_v}{\omega_- - E_\bfk \pm i \eta} - \frac{u_c \bar{u}_v v_c \bar{v}_v}{\omega_- + E_\bfk \pm i \eta} \biggr\} \\
\label{eq:62}
  \Bigl(\tilde{\Pi}_{\text{el},\bfq}^{R/A}(\omega)\Bigr)_{21} &=  \bar{g}^2 \sum_{\bfk} \biggl\{ \frac{\bar{u}_c u_v \bar{v}_c v_v}{\omega_+ - E_\bfk \pm i \eta} - \frac{\bar{u}_c u_v \bar{v}_c v_v}{\omega_+ + E_\bfk \pm i \eta} \biggr\} \\
\label{eq:63}
 \Bigl(\Pi_{\text{el},\bfq}^{R/A}(\omega)\Bigr)_{22} &= |g|^2 \sum_{\bfk} \biggl\{ \frac{|v_c|^2 |u_v|^2}{\omega_+ - E_\bfk \pm i \eta} - \frac{|u_c|^2 |v_v|^2}{\omega_+ + E_\bfk \pm i \eta} \biggr\} \,,
\end{align}
and the Keldysh components are given by
\begin{align}
  \label{eq:64}
  \Bigl( \Pi^K_{\text{el},\bfq}(\omega) \Bigr)_{11} &= - 2 \pi i |g|^2 \sum_{\bfk} \Bigl\{ |u_c|^2 |v_v|^2 \delta(\omega_- - E_\bfk) \nonumber \\ & \quad - |v_c|^2 |u_v|^2 \delta(\omega_- + E_\bfk) \Bigr\} \\
\label{eq:65}
  \Bigl( \Pi^K_{\text{el},\bfq}(\omega) \Bigr)_{12} &= - 2 \pi i g^2 \sum_{\bfk}  u_c v_c \bar{u}_v \bar{v}_v \nonumber \\ & \quad \times \Bigl\{ \delta(\omega_- - E_\bfk) - \delta(\omega_- + E_\bfk) \Bigr\}\\
\label{eq:66}
 \Bigl( \Pi^K_{\text{el},\bfq}(\omega) \Bigr)_{21} &= - 2 \pi i \bar{g}^2 \sum_{\bfk}  \bar{u}_c \bar{v}_c u_v v_v \nonumber \\ & \quad \times  \Bigl\{ \delta(\omega_+ - E_\bfk) - \delta(\omega_+ + E_\bfk) \Bigr\}\\
\label{eq:67}
  \Bigl( \Pi^K_{\text{el},\bfq}(\omega) \Bigr)_{22} &= - 2 \pi i |g|^2 \sum_{\bfk} \Bigl\{ |v_c|^2 |u_v|^2 \delta(\omega_+ - E_\bfk) \nonumber \\ & \quad - |u_c|^2 |v_v|^2 \delta(\omega_+ + E_\bfk) \Bigr\} \,.
\end{align}
Here, we have used the abbreviations
\begin{align}
  \label{eq:68}
  E_\bfk &= E_{c\bfk} + E_{v\bfk} \\
\label{eq:69}
  \omega_\pm &= \omega \pm (\mu_{c} - \mu_v) = \omega \pm e V_0 \,.
\end{align}
Note that we have neglected corrections to the quasi-particle energies due to the photon momenta and used $E_{\alpha \bfk + \bfq} \approx E_{\alpha \bfk}$. As a result, the photon self-energies are in fact independent of the photon momentum $\bfq$.

\subsection{Dressed photon propagator}
\label{sec:dress-phot-prop}
We obtain the dressed photon propagator $D_\bfq(t,t')$ by inverting the Dyson equation~\eqref{eq:55} up to the second order in the photon-electron coupling $g$:
\begin{align}
  \label{eq:59}
  D_\bfq(t,t') &= \tilde{D}_{0,\bfq}(t,t') \nonumber \\ & \quad + \int_{t_1, t_2} \tilde{D}_{0\bfq}(t - t_1) \Pi_{\text{el},\bfq}(t_1, t_2) \tilde{D}_{0,\bfq}(t_2,t') \,.
\end{align}
Note that we take the bath induced self-energy $\Pi_{\text{bath}}(t-t')$ fully into account by using the bath dressed photon propagator $\tilde{D}_{0,\bfq}$ (see Eq.~\eqref{eq:32}) instead of the bare one $D_{0,\bfq}$. When converting the integrals over time into integrals over frequency using Fourier transforms, one must be careful to take the relation in Eq.~\eqref{eq:58} into account.

Terms in Eq.~\eqref{eq:59} that involve normal components of the self-energy, which are functions of the difference of the two time arguments and appear in the diagonal entries of Eq.~\eqref{eq:56}, are of the form
\begin{align}
  \label{eq:70}
  \int_{t_1, t_2} &f(t,t_1) \Bigl( \Pi_{\text{el},\bfq}(t_1 - t_2)\Bigr)_{\bar{b} b/ b \bar{b}} h(t_2, t') \nonumber \\ & = \int_{\omega} e^{- i \omega (t - t')} f(\omega) \Bigl( \Pi_{\text{el},\bfq}(\omega)\Bigr)_{\bar{b} b/ b \bar{b}} h(\omega) \,,
  % \Bigl( D_\bfq(t,t') \Bigr)_{ii}= \Bigl( \tilde{D}_{0,\bfq}(t,t') \Bigr)_{ij} + \int_{\omega} e^{- i \omega(t-t')} \Bigl( \tilde{D}_{0,\bfq}(\omega) \Bigr)_{ik} \Bigl(\Pi_{\text{el}}(\omega) \Bigr)_{kk}   \Bigl( \tilde{D}_{0,\bfq}(\omega) \Bigr)_{kj}
\end{align}
where $\int_\omega = \int_{-\infty}^\infty \frac{d\omega}{2 \pi}$. In contrast, terms which contain anomalous components of the self-energy are of the form
\begin{align}
  \label{eq:71}
  &\int_{t_1, t_2} f(t,t_1) e^{\pm i 2 e V_0 t_2} \Bigl( \tilde{\Pi}_{\text{el},\bfq}(t_1 - t_2)\Bigr)_{\bar{b} \bar{b}/ b b} h(t_2, t') \nonumber \\ & = e^{\pm i 2 e V_0 t'} \int_{\omega} e^{- i \omega(t-t') } f(\omega) \Bigl( \tilde{\Pi}_{\text{el},\bfq}(\omega)\Bigr)_{\bar{b} b/ b \bar{b}} h(\omega\pm 2 e V_0) \,.
\end{align}
\subsection{Photon expectation values}
\label{sec:phot-expect-valu}
The anomalous photon expectation value $\av{\tilde{b}_{\bfq\lambda}(t) \tilde{b}_{-\bfq \lambda} (t)}$ where $\tilde{b}_{\bfq \lambda}(t) = b_{\bfq \lambda}(t) e^{i \omega_\bfq t}$ determines the degree of squeezing of the produced light via Eq.~(9) of the main text.
It is given by an anomalous component of the Keldysh Green's function at equal times $D^K_\bfq(t,t)$, which is calculated from Eq.~\eqref{eq:59}. At zero temperature, one finds
\begin{align}
  \label{eq:72}
  \av{\tilde{b}_{\bfq \lambda}(t) \tilde{b}_{-\bfq \lambda}(t)} &= i \Bigl( D^K_{\bfq}(t,t) \Bigr)_{12} e^{2 i \omega_\bfq t} \nonumber \\ &= \sum_{\bfk} \frac{2 g^2 u_{c \bfk} v_{c \bfk} \bar{u}_{v \bfk} \bar{v}_{v \bfk} e^{2 i (\omega_\bfq - 2 e V_0)t} }{ (\nu_\bfq - i \eta) (\nu_\bfq + E_\bfk - i \eta)} \,,
\end{align}
where $\nu_\bfq = \omega_\bfq - e V_0$ is the photon momentum measured relative to the applied voltage and $E_\bfk = E_{c \bfk} + E_{v \bfk}$. This is the result in Eq.~(12) of the main text.

The expectation value in Eq.~\eqref{eq:72} can also be analytically obtained in the presence of a thermal background of photons, or any other photon distribution function of the bath which enters via $\Pi^K_{\text{bath}}(\omega)$ in Eq.~\eqref{eq:50}. Since the analytical expression is rather lengthy, we choose to present a plot of the resulting expectation value $\av{\tilde{b}_{\bfq \lambda} \tilde{b}_{-\bfq \lambda}}$ (at $t=0$) for a thermal distribution of photons in Fig.~2c of the main text.

\subsection{Summation over electronic momenta}
\label{sec:summ-over-electr}
The summation over electronic momenta in Eq.~\eqref{eq:72} can be done analytically for the simplifying assumption of parabolic band dispersions
\begin{align}
  \label{eq:73}
  \varepsilon_{c \bfk} &= \frac{\bfk^2}{2 m^*_c} + \frac{D}{2} \\
\label{eq:74}
  \varepsilon_{v \bfk} &= -\frac{\bfk^2}{2 m^*_v} - \frac{D}{2}
\end{align}
with effective masses $m^*_c$, $m^*_v$ and band gap $D$. For a symmetric choice of (momentum independent) gap amplitudes $|\Delta_c| = |\Delta_v| \equiv |\Delta|$, effective masses $m_c^* = m_v^*$ and chemical potentials $\mu_c = D/2 + \delta$, $\mu_v = - D/2 - \delta = - \mu_c$, we find that $\xi_{c\bfk} =  \varepsilon_{c\bfk} - \mu_c = - (\varepsilon_{v \bfk} - \mu_v) = - \xi_{v\bfk}$. Here, $\delta$ refers to the location of the (quasi-) Fermi energies in the valence and conduction band due to doping and the applied strong forward bias~\cite{SalehTeich_Photonics}. Our choice of parameters is depicted in Fig.~\ref{fig:1}.
\begin{figure}[t]
  \centering
  \includegraphics[width=.6\linewidth]{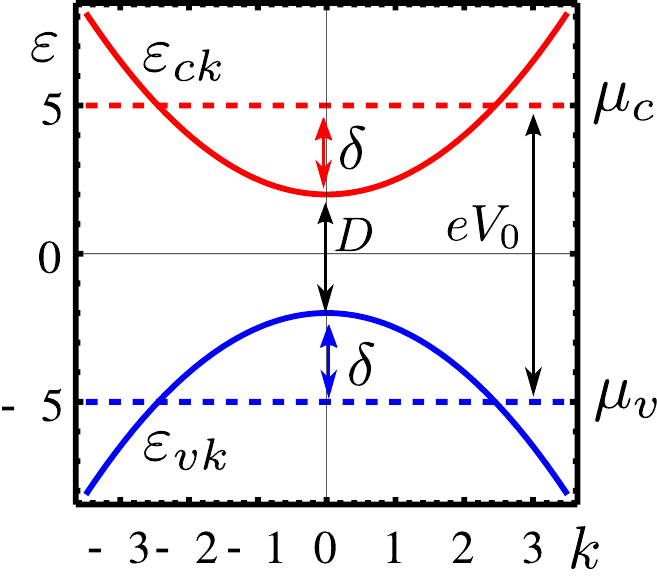}
  \caption{Schematic view of the band energy conventions with conduction and valence band dispersion $\varepsilon_{c\bfk}$ and $\varepsilon_{v \bfk}$, band gap $D$ and chemical potentials $\mu_c$ and $\mu_v$. For simplicity, we sketch the one-dimensional case. The location of the (quasi-) Fermi energies in the two bands are denoted by $\delta$ and the applied voltage by $e V_0 = \mu_c - \mu_v$.}
  \label{fig:1}
\end{figure}
It follows that
\begin{align}
  \label{eq:75}
  \int d^dk f(\xi_{c\bfk}, \xi_{v\bfk}) &= \int_{-\delta}^\infty d \xi \rho_c(\xi + \frac{D}{2} + \delta ) f(\xi, -\xi)
\end{align}
with the conduction band density of states
\begin{align}
  \label{eq:76}
  \rho_c(\epsilon) &= \theta\Bigl(\epsilon - \frac{D}{2} \Bigr) \frac{V (m_c^*)^{3/2}}{\sqrt{2} \pi^2 \hbar^3}  \sqrt{\epsilon - \frac{D}{2}  } 
%& \text{for $d=3$ } \\
%\frac{m_c^*}{2 \pi} & \text{for $d=2$. } \end{cases}
\end{align}
The sum of Bogoliubov energies takes the simple form
\begin{align}
  \label{eq:77}
  E(\xi_\bfk) &= E_{c\bfk} + E_{v\bfk} = 2 \sqrt{\xi_{\bfk}^2 + \Delta^2} \,.
\end{align}
We can thus write Eq.~\eqref{eq:72} in the form
\begin{align}
  \label{eq:78}
  \av{\tilde{b}_{\bfq \lambda}(t) &\tilde{b}_{-\bfq \lambda}(t)} = \frac{2 g^2 |\Delta_c| |\Delta_v| e^{i (\varphi_c - \varphi_v)}}{\nu_\bfq - i \eta}  \nonumber \\ & \times \sum_{\bfk} \frac{1}{ 4 E_{c\bfk} E_{v\bfk} [\nu_\bfq + E_{c\bfk} + E_{v\bfk} - i \eta]} \\
\label{eq:79}
&= \frac{ g^2 |\Delta_c| |\Delta_v| e^{i (\varphi_c - \varphi_v)}}{2 (\nu_\bfq - i \eta)} H(\nu_\bfq, \Delta, \eta) \,,
\end{align}
where we have defined the function
\begin{align}
  \label{eq:80}
  H(\nu_\bfq)  &= \int_{-\delta}^\infty d\xi \frac{ \rho_c(\xi + \frac{D}{2} + \delta) }{ (\xi^2 + \Delta^2) [ \nu_\bfq + 2 \sqrt{\xi^2 + \Delta^2} - i \eta]} \,.
\end{align}
Changing variables from $\xi$ to $E(\xi)$ (see Eq.~\eqref{eq:77}) and using that
\begin{align}
\label{eq:81}
  \frac{d \xi}{d E} &= \frac{|E|}{ 2 \sqrt{E^2 - 4 \Delta^2} }
\end{align}
we obtain
\begin{align}
\label{eq:82}
  H(\nu_\bfq) &= \int_{2 \Delta}^{2 \sqrt{\delta + \Delta^2} } dE  \; h(E) + \int_{2 \Delta}^\infty dE  \; h(E)
\end{align}
with
\begin{align}
\label{eq:83}
  h(E) = \frac{|E|}{ 2 \sqrt{E^2 - 4 \Delta^2} } \frac{\rho_c}{ \frac14 E^2 ( \nu_\bfq + E - i \eta)} \,.
\end{align}
Assuming that $\delta \gg 2 \Delta$, we can also extend the upper limit of the integral over the states below the Fermi level to infinity. We take the density of states $\rho_c$ to be approximately constant and equal to its value at the Fermi surface to obtain
\begin{align}
  \label{eq:84}
  H(\nu_\bfq) &= 2 \rho_c \Bigl(\frac{D}{2} + \delta\Bigr)  \nonumber \\ & \times  \int_{2 \Delta}^{\infty } d E \frac{ 2}{ E \sqrt{E - 2 \Delta} \sqrt{E + 2 \Delta} ( \nu_\bfq + E - i \eta) } \,.
\end{align}
The integral is performed using dimensionless variables $x = E/2 \Delta$, $\tilde{\nu}_\bfq = \nu_\bfq/2 \Delta$ and $\tilde{\eta} = \eta/2 \Delta$ and one finds
\begin{align}
  \label{eq:85}
  &H(\nu_\bfq) = \frac{\rho_c}{ \Delta^2} \int_1^{\infty} \frac{dx}{x \sqrt{x-1} \sqrt{x+1} (x + \tilde{\nu}_\bfq - i \tilde{\eta})} \\
\label{eq:86}
  &= \frac{\rho_c}{\Delta^2} \frac{i \pi \bigl[ -1 + \sqrt{1 + (i \tilde{\nu}_\bfq + \tilde{\eta})^2} \bigr] + 2 \arcsinh [ i \tilde{\nu}_\bfq + \tilde{\eta}] } { 2 (i \tilde{\nu})_\bfq + \tilde{\eta}) \sqrt{ 1 + (i \tilde{\nu}_\bfq + \tilde{\eta})^2} } \,.
\end{align}
Inserting this expression into Eq.~\eqref{eq:79}, we find the result of Eq.~(13) in the main text
\begin{align}
  \label{eq:87}
  \av{\tilde b_{\bfq \lambda}(t) &\tilde b_{-\bfq \lambda}(t)} = g^2 \rho_c |\Delta|^2 e^{i \Delta \varphi} e^{2i(\omega_\bfq - e V_0)t}  \\ &\times \frac{ 2 \arcsin(\tilde{\nu}_\bfq - i\tilde{\eta}) + \pi [ -1 + \sqrt{1 - (\tilde{\nu}_\bfq -i \tilde{\eta})^2} ] } {(\nu_\bfq - i \eta)^2 \sqrt{ 4 |\Delta|^2 - (\nu_{\bf q} - i \eta)^2} } \nonumber \,,
\end{align}

%\bibliography{Biblio}

\end{document}